\documentclass{article}

\usepackage{amssymb,amsmath}
\usepackage{fullpage}
\usepackage{graphicx}
\usepackage{amsmath}		
\usepackage[margin=1.0in]{geometry}
\usepackage{setspace}
\usepackage{color}
\usepackage{fancyhdr}
\usepackage{collcell}
\usepackage{datatool}
\usepackage{environ}
\usepackage{latexsym}
\usepackage{amssymb}
\usepackage{epsfig,amsmath,graphics}
\usepackage{epstopdf}
\usepackage{verbatim}
\usepackage{wasysym}
\usepackage{feynmp-auto}
\usepackage{authblk}
\usepackage{xcolor}
\usepackage{enumitem}
\usepackage[utf8]{inputenc}
\usepackage{slashed}
\usepackage{cite}

\usepackage{empheq}

    \newlength\fsep
    \newsavebox\widebox

\newcommand{\et}{\tilde{\epsilon}}

\usepackage[skip=0pt]{caption}

\begin{document}
\title{A Causal Framework for Non-Linear Quantum Mechanics}
\author{David E.~Kaplan}
\author{Surjeet Rajendran}
\affil{\small Department of Physics \& Astronomy, The Johns Hopkins University, Baltimore, MD  21218, USA}

\date{\today}
\maketitle

\begin{abstract}
  We add non-linear and state-dependent terms to quantum field theory.  We show that the resulting low-energy theory, non-linear quantum mechanics, is causal, preserves probability and permits a consistent description of the process of measurement. We explore the consequences of such terms and show that non-linear quantum effects can be observed in macroscopic systems even in the presence of de-coherence. We find that current experimental bounds on these non-linearities are weak and propose several experimental methods to significantly probe these effects. The locally exploitable effects of these non-linearities have enormous technological implications. For example, they would allow large scale parallelization of  computing (in fact, any other effort) and enable quantum sensing beyond the standard quantum limit. We also expose a fundamental vulnerability of any non-linear modification of quantum mechanics - these modifications are highly sensitive to cosmic history and their locally exploitable  effects can dynamically disappear if the observed universe has a tiny overlap with the overall quantum state of the universe, as is predicted in conventional inflationary cosmology.  We identify observables that should persist in this case and discuss opportunities to detect them in cosmic ray experiments, tests of strong field general relativity  and current probes of the equation of state of the universe. Non-linear quantum mechanics also enables novel gravitational phenomena and may open  new directions to solve the black hole information problem and to uncover the theory underlying quantum field theory and gravitation. 
\end{abstract}

\section{Introduction}

Quantum Mechanics is the bedrock of physics. Its seemingly ad-hoc and phenomenologically derived axioms have proven to be remarkably resistant to parametrized deviation. Given its central importance to physics, there is a clear need to develop a consistent framework to test deviations from these axioms. But what parts of quantum mechanics can we modify? Quantum theory rests on two elements.  First, it unavoidably introduces an element of probability into physics due to the act of measurement. Second, it requires time evolution to be linear. 

The unease triggered by the fundamental role of probability has prompted several efforts to modify the theory to make its predictions absolutely deterministic. In our view, the elimination of probability as an integral element of describing the outcome of physical processes does not seem likely. This is due to the physical fact that a finite system is only allowed to have a finite set of energies, even though its constituents possess continuous observables and symmetries. These aspects of physical systems are placed in conflict when one tries to impose the concept of a deterministic measurement of {\it all} the observables of the system. As an example, consider the ground state of the hydrogen atom. The electron in the atom has continuous observable properties, for example, its position relative to the proton.  Moreover, the system is also endowed with rotational symmetry. If the electron could exist in a state with a deterministic position and energy, application of the rotational symmetry would generate an infinite set of states that all have the same energy but with different electron positions -- a continuous degeneracy. This would then contradict the assumption that the system only has a single ground state (and in general, a finite set of energy eigenstates below some energy). Probability allows quantum mechanics to retain a finite set of energy states  with the existence of continuous symmetries and observables by sacrificing deterministic measurement. The Bell inequalities \cite{Bell:1964kc}, the SSC \cite{Mermin:1998cg} and Kochen-Specker theorems \cite{Kochen:1968zz} are mathematical illustrations of this conflict. 

Motivated by the above, in this paper we pursue the possibility of modifying the second element of quantum mechanics, namely, linear time evolution. There have been several attempts to introduce non-linear evolution into quantum mechanics  \cite{Weinberg:1989us} and they generically suffer from problems of causality
\cite{Polchinski:1990py, GISIN19901}.

Additionally, non-linear evolution requires a careful treatment of the concept of measurement.  Our goal in this paper is to provide a causal non-linear modification of quantum mechanics with a consistent description of measurement. The key ingredients that enable such a description are the following. First, the modification proposed by us breaks the basis independence of the Schr\"{o}dinger equation and is explicitly written in terms of the position basis. The basis independence of the Schr\"{o}dinger equation, while mathematically pleasing, is at odds with the physical fact that interactions are local, resulting in the position basis being a preferred basis to describe the world.  The position basis can also manifestly enforce causality - precisely because causality is fundamentally a consequence of the local nature of interactions. Second, we will not describe measurement as a phenomenon that is somehow distinct from any of the other interactions in the theory. Instead, we will adopt the view that measurement arises as the result of an interaction between a measuring device and a quantum system, with the interaction being described by the time evolution equations that govern any other interaction in the theory.\footnote{This view parallels the treatment of measurement in the ``many-worlds'' interpretation of quantum mechanics - one cannot implement, for example, the ``Copenhagen interpretation'' of measurement with its ad-hoc rule causing the ``collapse'' of the wave-function since such a collapse causes violence to non-linear terms, and is anyway itself a modification of quantum mechanics.} Using this concept of measurement,  while causally consistent, we will find it to be fundamentally in conflict with the notion of a measuring device whose measurements are accurate and repeatable. Non-linear evolution continually affects the evolution of a  systems in a state-dependent way and  introduces a fundamental notion of error into all measurements. 

The specific modification that we investigate is described in detail in the next section. The basic procedure that we follow is to take the Hamiltonian of a quantum field theory and include additional state-dependent terms in it. These additional terms are obtained by taking existing interactions in the theory and replacing the bosonic (or fermion bi-linear) field operators  by their expectation values in the full quantum state.  Thus it can be implemented in any interacting theory.  This additional (state-dependent) expectation value manifests as a background field and adds an element of classical time-evolution to the theory. In low-energy single-particle quantum mechanics, it manifests itself as a (causal) non-linear term in the wave function's time evolution.  Our goal in this paper is to establish the viability of this framework and investigate the phenomenological, experimental and cosmological implications of this modification.


This paper is organized as follows. In Section \ref{sec:framework}, we develop a transparent framework to show that the modification can be implemented in a causal and gauge invariant manner. We show  that this modification allows for the existence of stationary states  (Section \ref{subsec:stationary}), an essential element of quantum systems and we develop a consistent notion of measurement (Section \ref{subsec:measurement}) and derive its consequences. In Section \ref{sec:analysis}, we analyze the consequences of this framework. We show that non-linear quantum effects can be observed even in the presence of decoherence. However, the human ability to exploit these non-linear effects are subject to our cosmic history (Section \ref{subsec:cosmo}) wherein these effects can dynamically disappear if our macroscopic universe was produced as a result of quantum fluctuations as is the case in canonical inflationary theories. We highlight that this aspect of non-linear quantum mechanics is not unique to our modification but is rather a general feature of any causal non-linear modification. Non-linear quantum mechanics can lead to unusual gravitational and causal phenomena and we discuss these in Section \ref{subsec:quantumwormholes}. In Section \ref{sec:pheno} we analyze current experimental limits on such non-linearities  and find them to be surprisingly weak. In Section \ref{sec:experiments},  we then propose experimental protocols to test these non-linearities for different classes of cosmic history - one where our universe originated and evolved as a result of deterministic classical processes and another where it was the result of quantum processes. There have been prior investigations of this class of modifications \cite{Kibble:1978vm, Stamp:2015vxa, Barvinsky:2018tsw, Carney:2018ofe, Barvinsky:2020pvw, Wilson-Gerow:2020moi} as well as more general investigations of non-linear quantum mechanics \cite{Weinberg:1989us, Polchinski:1990py}. We discuss the ways in which our efforts have advanced these investigations in Section \ref{sec:comparison}. We then conclude in Section \ref{sec:concl}.  

\section{The Framework} 
\label{sec:framework}
We begin by describing  non-linear quantum  evolution in field theory prior to describing these effects on single particle quantum mechanics (see Section \ref{subsec:singleparticle}). This order, while unconventional, nevertheless adds clarity and makes it clear the modification is derived from a local theory. Field theory is an accurate description of nature. It naturally describes multi-particle separated systems and by explicitly recognizing the special nature of position it enforces causality. It thus transparently addresses issues that have confronted prior non-linear modifications of quantum mechanics.  We first consider flat space before generalizing this description to include gravitation. Throughout this paper, we work in the Schr\"{o}dinger picture where the quantum states $|\chi\left(t\right)\rangle$ are time dependent while the operators are time independent.  

In the Schr\"{o}dinger picture of quantum field theory, the states $|\chi\left(t\right)\rangle$ time evolve as per the Schr\"{o}dinger equation: 

\begin{equation}
    i \frac{\partial |\chi\left(t\right) \rangle}{\partial t} = H |\chi\left(t\right)\rangle
    \label{QFTSchrod}
\end{equation}
where $H = \int d^3 \mathcal{H}\left({\bf x}\right)$ is the Hamiltonian of the theory and $\mathcal{H}\left({\bf x}\right)$ is the Hamiltonian density constructed from the time independent field operators. Notice that this Schr\"{o}dinger equation for $|\chi\left(t\right) \rangle$ can be obtained from the following action:

\begin{equation}
    S_{QFT} = \int dt \, \frac{i}{2}\left( \langle \dot{\chi} | \chi \rangle - \langle \chi | \dot{\chi} \rangle \right)  + \langle \chi | H | \chi \rangle 
    \label{QFTAction}
\end{equation}
The Schr\"{o}dinger equation \eqref{QFTSchrod} can be obtained as an extremum of the action \eqref{QFTAction} under variations of the path $|\chi\left(t\right)\rangle$. Consider, for example, the Yukawa theory with the interaction $ y\, \phi({\bf x}) \bar{\Psi}({\bf x}) \Psi({\bf x})$, depicting a scalar $\phi$ interacting with a fermion $\Psi$ with Yukawa coupling $y$. Notice that the action \eqref{QFTAction} of this theory contains the term:  
\begin{equation}
    S_{QFT} \supset \int dt d^3{\bf x} \,  y \langle \chi\left(t\right) | \phi\left({\bf x}\right) \bar{\Psi}\left({\bf x}\right) \Psi\left({\bf x}\right) | \chi\left(t\right)\rangle
    \label{QFTYukawa}
\end{equation}
We incorporate non-linear quantum mechanical evolution for this Yukawa theory by constructing the non-linear action:
\begin{equation}
S_{NL} = S_{QFT} + \epsilon \int dt d^3{\bf x} \, y {\langle \chi\left(t\right) | \phi\left({\bf x}\right) | \chi\left(t\right) \rangle \langle \chi\left(t\right) | \bar{\Psi}\left({\bf x}\right)\Psi\left({\bf x}\right) | \chi\left(t\right) \rangle}
\label{NLAction}
\end{equation}
where $\epsilon$ is the strength of the non-linearity\footnote{We could also divide by the (conserved) normalization of the state, $\langle\chi\left(t\right) |\chi\left(t\right)\rangle$, but we will set this equal to unity for notational simplicity, but keeping track of this normalization when discussing probabilities.}. The integrand of \eqref{Eqn:NLYukawa} is time reversal invariant, just like the integrand of \eqref{QFTAction}. This ensures that the resulting evolution  is unitary. Moreover, in the vacuum state, the correction vanishes and thus the evolution of the vacuum is Lorentz invariant. When the quantum state is not the vacuum, the non-linearity appears as a state-dependent in-medium effect. In section \ref{subsec:separated}, we show that these in-medium effects are causal. 


The time evolution of $|\chi\left(t\right)\rangle$ is obtained by computing the extrema of the action \eqref{NLAction} leading to time evolution of the form:
\begin{equation}
\label{Eqn:NLYukawa}
    i \frac{\partial |\chi\left(t\right)\rangle}{\partial t} = \left(H_L  + \epsilon \, y \,  \int d^3 {\bf x} {\left(\langle \chi\left(t\right) |\phi\left({\bf x}\right)|\chi\left(t\right) \rangle \bar{\Psi}\left({\bf x}\right) \Psi\left({\bf x}\right)  + \langle \chi\left(t\right) |\bar{\Psi}\left({\bf x}\right) \Psi\left({\bf x}\right) |\chi\left(t\right) \rangle \right) \phi\left({\bf x}\right)}
    \right) |\chi\left(t\right) \rangle
\end{equation}
where $H_L$ is the standard state independent ({\it i.e.} linear) Hamiltonian of the Yukawa theory. The non-linear action \eqref{NLAction} and the corresponding state dependent non-linear Hamiltonian of the theory are natural extensions of the corresponding linear theory obtained from the action \eqref{QFTYukawa}. In the above, we have included non-linear terms from both the expectation value of $\phi$ and the fermion bi-linear $\bar{\Psi}\Psi$. This is a consequence of \eqref{NLAction} which is a natural extension of \eqref{QFTAction} and as we will see in Section \ref{subsec:singleparticle} permits a natural way to construct conserved quantities in this theory, as we will discuss in Section \ref{eqn:stationary}. But, there is no requirement to include both expectation values as it is possible to construct non-linear quantum mechanical theories that only include the expectation value of $\phi$ (for example) and show in detail that energy (and other quantities) are conserved in such theories.

The quantization of this theory can be described perturbatively. At zeroth order, the theory is simply standard quantum field theory and one may use standard canonical procedures to define the quantum states of the theory. Non-linear corrections can be computed perturbatively and these only require knowledge of the lower order terms. Effectively, in this theory, the lower order terms contribute as background classical fields, permitting a straightforward analysis of the quantum evolution. 

The Hamiltonian evolution in \eqref{Eqn:NLYukawa} can also be described using the path integral formalism. The specific non-linear interaction in \eqref{NLAction} can be described perturbatively by computing the path integral with the effective Lagrangian: 

\begin{equation}
    \mathcal{L}_{eff} \supset y \phi \bar{\Psi} \Psi + \epsilon y {\langle \chi | \phi |\chi \rangle}
    \bar{\Psi} \Psi + \epsilon y \phi {\langle \chi| \bar{\Psi} \Psi | \chi \rangle}
\end{equation}
where we have included additional ``interaction'' terms in the Lagrangian with the state-dependent expectation values being treated as background classical fields (themselves evaluated at lower order in $\epsilon$) in the path integral.

With an eye towards the single particle theory, let us describe how the fermion $\Psi$ behaves in this theory. To first order in $\epsilon$, $\Psi$ responds to: 
\begin{equation}
     y \phi + y \epsilon  {\langle \chi | \phi | \chi \rangle }
\end{equation}
{\it i.e.}, both the quantum field $\phi$ and the background classical field created by the expectation value ${\langle \chi | \phi | \chi \rangle }$.  The expectation value  $\langle \chi |\phi\left(x\right)|\chi \rangle$ of $\phi$ in the state $|\chi\rangle$  is obtained from the full interacting theory. Evaluating this expectation value perturbatively in $y$ and to zeroth order in $\epsilon$: 

\begin{equation}
\label{eqn:Yukawa}
    \langle \chi  | \phi \left(x\right) | \chi \rangle = y\, \int d^4 x_1   \langle \chi  | \bar{\Psi} \left({\bf x_1}\right) \Psi \left({\bf x_1}\right)| \chi \rangle G_R \left(x;  x_1\right) + \dots
\end{equation}
where $G_R \left( x;   x_1\right)$ is the retarded Green's function from the space-time point $x_1$ to $x$ of the massless field $\phi$ and  $x_1$, for example, is the 4-vector $(t,{\bf x_1})$. The ellipsis includes the time-evolved initial condition for the $\phi$ field itself.  In the above, we use the renormalized expression for $\langle \chi| \bar{\Psi}  \Psi | \chi \rangle$ which removes the vacuum energy divergence in this expression. 

This construction can also be extended to gauge theories and gravitation. These theories are more easily described using the Lagrangian formalism rather than the Hamiltonian formalism and we will thus use the former. Consider a $U\left(1\right)$ gauge field $A_{\mu}$. Notice that under a gauge transformation $A_{\mu}\rightarrow A_{\mu} + \partial_{\mu}\alpha$, 

\begin{equation}
    B_{\mu} =  \frac{A_{\mu} + \epsilon  {\langle \chi | A_{\mu}|\chi \rangle}
    }{1 + \epsilon } 
\end{equation}
also transforms as $B_{\mu} \rightarrow B_{\mu} +\partial_{\mu}\alpha $. Using this fact, we can construct a gauge invariant non-linear quantum mechanical action for $A_{\mu}$ by taking the interaction term $eJ^{\mu}A_{\mu}$ of the theory and replacing $A_{\mu} \rightarrow B_{\mu}$. For example, in QED this leads to non-linear interactions of the form: 
\begin{equation}
  \mathcal{L}_{QED} \supset e   A_{\mu} \bar{\Psi}\gamma^{\mu}\Psi  \rightarrow   \mathcal{L}_{eff} \supset  e   B_{\mu} \bar{\Psi}\gamma^{\mu}\Psi  
  \label{QED1}
\end{equation}
Perturbatively, the effect is essentially an additional background electromagnetic field treated classically (no different than how we do QED at low energies).

Gauge invariant non-linear interactions can also be incorporated via the kinetic terms. For example, in QED: 
\begin{equation}
    \mathcal{L}_{QED} \supset  F_{\mu \nu}^2  \rightarrow \mathcal{L}_{eff} \supset   F_{\mu \nu}^2  + \epsilon \langle \chi | F_{\mu \nu} |\chi \rangle  F^{\mu \nu}
    \label{QED2}
\end{equation}
In general, the non-linear coefficients in $\eqref{QED1}$ and $\eqref{QED2}$ need not be the same. Moreover, the terms in $\eqref{QED1}$ need not be flavor universal. For simplicity, in this paper, we will set the non-linear term in $\eqref{QED2}$ to zero and take $\eqref{QED1}$ to be flavor universal. We will address non-flavor universal effects in vector-like and chiral theories in future work.

Much like the Yukawa case, in the quantum theory of the fermion $\Psi$, ${ \langle \chi | A_{\mu}|\chi \rangle}$ is  a classical background. 
After gauge fixing, one finds the leading order contribution to this expectation value to be (perturbatively):
\begin{equation}
\label{eqn:gauge}
 \langle \chi | A_{\mu} | \chi \rangle = e \, \int d^4 x_1  \langle \chi  | \bar{\Psi} \left({\bf x_1}\right)\gamma^{\nu} \Psi \left({\bf x_1}\right)| \chi \rangle G_{R\mu\nu} \left( x;  x_1\right) + \dots
\end{equation}
where $G_{R\mu\nu}$ is the retarded Green's function. The gauge invariance of the Lagrangian implies that we can still gauge fix $A_{\mu}$ and define its physical degrees of freedom. 

In principle, a similar construction can also be performed for non-Abelian gauge theories. However, the quantum consistency of the non-Abelian theory may imply additional constraints on the non-linear coefficients $\epsilon$ - we leave a detailed analysis of this consistency check for future work.

This non-linear construction can also be extended to gravitation. Similar to the case of $U\left(1\right)$ electromagnetism, one can observe the fact that $\tilde{g}_{\mu \nu}$ defined by: 

\begin{equation}
     \tilde{g}_{\mu \nu} = \frac{g_{\mu \nu } +  \epsilon {\langle \chi | g_{\mu \nu}| \chi \rangle}}{1 + \epsilon} 
    \label{GR1}
\end{equation}
has the same tensor properties as the metric $g_{\mu \nu}$. By writing the matter terms in the Lagrangian of General Relativity  using the metric $\tilde{g}$ instead of $g$, we obtain a  covariant way to incorporate non-linear quantum mechanics into General Relativity. Similar in spirit to \eqref{QED2}, it may also be possible to introduce non-linearities via the kinetic terms. We explore these possibilities in future work.

Super-position is a fundamental feature of quantum mechanics and the theory admits states that feature linear combinations of space-time metrics. Our formalism permits a well-defined way to include non-linearity into quantum gravity thus permitting, in principle, computable cosmological consequences of these non-linearities, something we discuss in part in the next section and more fully in future work. For now, we simply note that the elements of the Lagrangian are tensor fields on the manifold and they have a geometric meaning independent of any specific metric. There is thus no ambiguity in defining the action.

In the following subsections, we derive the  single particle Schr\"{o}dinger equation (\ref{subsec:singleparticle}) and show that this equation allows quantum states to have a conserved norm,  permitting a probabilistic interpretation of the theory. In \ref{subsec:stationary} we show that this framework allows quantum systems to have stationary states with well defined energies. We describe separated systems and show that the theory is causal (\ref{subsec:separated}), and then describe the meaning of and the effects on measurement in our modification (\ref{subsec:measurement}). 

\subsection{Single Particle Quantum Mechanics} 
\label{subsec:singleparticle}
The non-linear evolution proposed by us is interaction (and field) dependent. Thus, the single-particle Schr\"{o}dinger equation that results from this modification depends on the specific field theory. We describe these effects for a fermion $\Psi$ that interacts via a Yukawa coupling $y \phi \bar{\Psi} \Psi$ with a scalar $\phi$. The state of a single $\Psi$ quantum is described by its wave-function  $\Psi\left(t, {\bf x}\right)$ (where $x = (t,{\bf x})$ and three-vectors are in bold). Under the non-linearity, the time evolution of this wave-function is governed by the equation:  

\begin{equation}
    i \frac{\partial \Psi \left(t, {\bf x}\right)}{\partial t} =\left( H  + \epsilon\, y \int d^4 x_1 \Psi^{*}\left(t_1,{\bf x}_1\right) \Psi\left(t_1,{\bf x}_1\right) G_{R}\left( x; x_1\right)\right) \Psi\left(t, {\bf x}\right)
    \label{Eqn:NLS}
\end{equation}
where $H$ is usual Hermitian Hamiltonian of quantum mechanics in the position representation and $G_{R}$ is the relativistic retarded Green's function between the points $\left(t, {\bf x}\right)$ and  $\left(t_1,{\bf x}_1\right)$.   This equation governs the evolution of the position degrees of freedom of the state and we trace over internal quantum numbers (such as spin). Notice that this equation is invariant under the transformation $\Psi\left(t, {\bf x}\right) \rightarrow e^{i \theta} \Psi\left(t, {\bf x}\right) $. It thus admits a conserved probability current and norm, permitting a probabilistic interpretation of the wave function. 
 
The above equation has the same form as the Schr\"{o}dinger-Newton equation that is sometimes described as the non-linear Schr\"{o}dinger equation. But, the non-linear Schr\"{o}dinger equation is used to describe the linear quantum mechanical evolution of a particle in the mean field of a large number of particles all of whom are in the same quantum state. In \eqref{Eqn:NLS}, we are proposing a non-linear quantum evolution of a single particle state. As we will see below, this non-linear behavior leads to considerably different physical phenomena than the linear quantum evolution of a multi-particle system. 

There is a key mathematical difference between the Schr\"{o}dinger equation  and \eqref{Eqn:NLS}. Formally, this equation requires knowledge of the full past of $\Psi\left(x\right)$ in order to describe its future evolution as opposed to simply its value at a given time $t_0$. Alternately, one may specify the value of the wave-function at $t_0$ and specify an additional boundary condition to capture the effects of the past and then perform the non-linear evolution. The latter approach is physically motivated:  equation \eqref{Eqn:NLS} arises from a Yukawa interaction that allows the system to interact with itself via the sourced scalar field. The additional boundary condition is then the specification of the background value of this  scalar field at that specific time slice. Even though, in principle,  the evolution of $\Psi\left(x\right)$ requires the full knowledge of the past of $\Psi\left(x\right)$, in practise, the effects of distant events on its evolution are suppressed. This is because the influence of the past is felt via the Green's function in \eqref{Eqn:NLS} which naturally suppresses the effects of the past on the current evolution of the system. In this aspect, the time evolution of the quantum state $\Psi\left(x\right)$ is similar to the evolution of a two (or multi) particle state in an interacting quantum theory where the time evolution of the state does depend (in principle) on the full past of all the particles in the state. But, the effects of events in the past are suppressed precisely by these Green's functions. 
 
\subsection{Stationary States and Energy Conservation}
\label{subsec:stationary}
The existence of stationary states is one of the central consequences of quantum mechanics. It allows for the existence of ground states of quantum systems without which we cannot define a stable vacuum. Using perturbation theory and induction, it is straightforward to show that  non-linear evolution  permits the existence of stationary states. We will explicitly demonstrate this for single particle quantum mechanics. It is appropriate to use perturbative arguments since the non-linear corrections do not source instabilities or runaway potentials in the Hamiltonian, as long as the correction is small enough for a given potential (for shallow binding potentials, the existence of bound states can be used to constrain the non-linearity, see Section \ref{sec:pheno}). Similar arguments can also be applied to the field theory case where we can perturbatively describe single particle states. Note that this perturbative argument shows that if the linear quantum mechanical Hamiltonian was bounded from below, sufficiently small non-linear perturbations do not change that fact. If the linear quantum mechanical Hamiltonian was unbounded from below, perturbative non-linear effects are unlikely to alter that aspect of the spectrum.  


We now show the existence of stationary states in single particle quantum systems of the Yukawa theory described by \eqref{Eqn:NLS}. We will take the scalar field to be massless so it is relevant to the cases of the photon and graviton, though the inclusion of a mass is straight forward.  In this case, the Green's function that appears in \eqref{Eqn:NLS} would be that of a massless Klein-Gordon field, $G_R\left(t, {\bf x}; t_1, {\bf x}_1\right) = \delta\left( t_1 - t - |{\bf x}_1 - {\bf x}|\right)/\left(4 \pi |{\bf x}_1 - {\bf x}|\right) $.

By a single-particle stationary state, we mean a state that has a wave function whose sole time dependence is an overall phase, $\Psi\left(t, {\bf x}\right) = e^{- i E t}\Psi\left(0, {\bf x} \right)$, and thus satisfies the equation: 
\begin{equation}
\label{eqn:stationary}
   \left[ H +  \epsilon \underbrace{y {\int d^3 {\bf x_1} \frac{\Psi^{*}\left(0, {\bf x}_1 \right)\Psi\left(0, {\bf x}_1\right)}{4 \pi |{\bf x}_1 - {\bf x}|}}}_{V_{NL}}\right] \Psi\left(t, {\bf x}\right) = E\, \Psi\left(t, {\bf x}\right)
\end{equation}
Having done the time integral in the non-linear component, we see this term is a state-dependent, but time-independent, correction to the effective Hamiltonian. Thus, for Hamiltonians with known spectra (such as the hydrogen atom), one can solve for stationary states perturbatively in $\epsilon$ by expanding $\Psi$, $E$, and $V_{NL}$ as

\begin{eqnarray*}
    \Psi& = &\Psi_0  + \epsilon \Psi_1  + \epsilon^2 \Psi_2 + \cdots \\ \\
    E& = &E_0  + \epsilon E_1  + \epsilon^2 E_2 + \cdots \\ \\
    V_{NL}\left[\Psi,\Psi^*\right]& = &V_{NL}\left[\Psi_0,\Psi_0^*\right] + \epsilon \int_{z_1}\left( \Psi_1\left(0,{\bf z}_1\right) \left(\frac{\delta V_{NL}}{\delta \Psi\left(0,{\bf z}_1\right)}\right)_{\Psi=\Psi_0} + h.c.\right) + \cdots
\end{eqnarray*}
and solving equation \eqref{eqn:stationary} order by order.  Note that the potential term $V_{NL}$ is already at order $\epsilon$. Thus, for example, the $k^{th}$ order equation will only involve potential terms with $\Psi_i$'s for $i<k$, and thus the equations will be linear and, in general, solvable.

%
%

The existence of stationary states has  key implications for the stability of the theory - it ensures that the theory has stable ground states and prevents runaway violations of energy conservation. To see this, observe that the probability densities of stationary states are independent of time.  The non-linear terms that appear in their time evolution are time independent. This time translation invariance implies that the energies of these states are conserved. This is reassuring as these are the only states of the theory that have a well defined energy. A general state that is a linear combination of these stationary states will have  time dependent probability densities and the non-linear terms that govern their evolution are time dependent. The energy of these states, as measured by the non-linear Hamiltonian, is not conserved.  This is not surprising since these states do not actually have a well defined energy. However, even though the energy of the state is not conserved, one does not expect disastrous runaway violations of energy conservation since the system possesses a ground state. In linear quantum mechanics this is proven by constructing a positive definite conserved energy functional - namely, the expectation value of the stress energy tensor. This stress tensor is obtained from a time translation invariant Lagrangian. 

One can show that an actual conserved energy can be constructed for this non-linear modification. Instead of the one particle equation, we use the underlying field theory where  energy conservation is trivial to show via  Noether's Theorem. This is because the integrand in \eqref{NLAction} that is used to obtain the theory is time translation invariant. It thus possesses a conserved energy. This conserved energy  depends on the normalized expectation value of $\phi$.  Thus some of the conserved energy can be seen to be stored in the background field and in general (as we will see examples later) can move between dynamic fields and the background.


\subsection{Separated  Systems and Causality}
\label{subsec:separated}

A key obstacle in prior attempts \cite{Weinberg:1989us} to introduce non-linear evolution in quantum mechanics has been the ability to describe separated but entangled systems while preserving causality. The field theoretic underpinning of our modification is  local and thus should be causal.  It also enables us to handle these issues in a transparent manner in the low-energy theory. 
It is instructive to demonstrate in our framework this causal evolution explicitly for a two particle system.  This demonstration will also illustrate some key features of the non-linear but unitary nature of the time evolution. For definiteness, we again illustrate this in the case of the Yukawa theory in flat space but our results apply more generally.  We direct the reader to Section \ref{subsec:quantumwormholes} for unusual causal phenomena that are possible in curved space-times. 

Suppose we have two particles  described by the co-ordinates ${\bf x}$ and ${\bf y}$ respectively. Following the field theory insertion of \eqref{eqn:Yukawa}, one can generate a clear prescription for multi-particle states, and it is easy to show that the time evolution of the two-particle wave function $\Psi\left({\bf x}, {\bf y};t\right)$ is described by:
\begin{eqnarray}
 \label{eqn:separated}
i \frac{\partial }{\partial t} \Psi\left({\bf x},{\bf y} ; t\right) &=& \left(H  + {y \,\epsilon}  \int  d^3 {\bf x}_1 d^3 {\bf y}_1 d\tau \, \left| \Psi\left({\bf x}_1, {\bf y}_1 ; \tau \right) \right|^2 \right. \\ \nonumber && \left. \times \left[ G_R\left(t,  {\bf x}; \tau, {\bf x}_1\right) +   G_R\left(t,  {\bf y}; \tau, {\bf x}_1\right)+  G_R\left( t, {\bf x}; \tau, {\bf y}_1\right)  + 
   G_R\left(t,  {\bf y}; \tau, {\bf y}_1 \right)\,\right] \phantom{\int}\!\!\!\!\!\!\!\! \right)   \Psi\left({\bf x},{\bf y} ; t\right)
\end{eqnarray}
We now show that this prescription preserves causality when we have a well separated entangled state. The second and third Green's functions serve to causally connect the sub-systems at ${\bf x}$ and ${\bf y}$ via non-linear evolution - these terms are thus not a threat to causality. Causal problems could potentially arise from the first and last terms. For example, suppose some local operation is performed on the particle in the ${\bf x}$ region. This would change its wave-function. If the change to this wave-function changes the time evolution of the particle in the ${\bf y}$ region solely through the fourth Green's function in \eqref{eqn:separated} ({\it i.e.} without the aid of the second causal Green's function), we would have violated causality. We will show that this is not the case. 

The term containing the fourth Green's function in \eqref{eqn:separated}: 
\begin{equation}
    \int d^3{\bf x_1} d^3 {\bf y_1} d\tau | \Psi\left({\bf x_1}, {\bf y_1}; \tau \right)|^2 G_R\left(t, {\bf y}; \tau, {\bf y_1}\right)
    \label{eqn:trouble}
\end{equation}
would produce causality violation if a local operation on ${\bf x}$ changed the integral over the ${\bf x_1}$ co-ordinates: causality would be violated because the change would instantly affect the evolution of the particle at {\bf y}. In the absence of the second term in \eqref{eqn:separated}, we will see that the full non-linear time evolution due to local operations on ${\bf x}$ can be represented by unitary operators of the form $U = U_1\left({\bf x_1}\right)\otimes U_2\left({\bf y_1}\right)$ and the time evolution results in $\Psi\left({\bf x}, {\bf y}, t\right) \rightarrow U\left[ \Psi\left({\bf x}, {\bf y}, t\right)\right]$ with separate unitary transformations on the separated systems. When this is the case, the integral over ${\bf x_1}$ in \eqref{eqn:trouble} does not change and causality is preserved. 

The unitary, factorized form of this time evolution can be proven in our non-linear case using time dependent perturbation theory. At zeroth order, the time evolution of the system is given by linear quantum mechanics. In this case, a local operation on {\bf x} is indeed represented by a factorized unitary operator on the full system.  The first order correction from the non-linear dynamics is obtained by using the zeroth order wave-function $\Psi_0$. In the absence of the second and third terms in \eqref{eqn:separated}, assuming the systems are causally disconnected, this first order correction is operationally no different than computing the first order correction due to the time dependent potential 

\begin{equation}
    V_{1}\left(t, {\bf x}\right) = \epsilon \int d^3 {\bf x_1} d^3 {\bf y_1} d\tau \Psi_{0}^{*} \Psi_0 G_R\left(t, {\bf x}; \tau, {\bf x_1}\right)
    \label{eqn:addition}
\end{equation}
That is, the non-linear correction is equivalent to computing the effect of a potential that solely depends on ${\bf x}$ in linear quantum mechanics. But, in the latter case, since we are now dealing with linear quantum evolution of two separated systems, the time evolution is indeed represented by a factorized unitary operator. Note that in \eqref{eqn:addition} we have only written down the non-linear correction to ${\bf x}$. There is a similar {\it additive} term that acts on ${\bf y}$ arising from the fourth Green's function in \eqref{eqn:separated}. The effective correction is thus of the form $V_1\left({\bf x}\right) + \tilde{V}_{1}\left({\bf y}\right)$. Due to its additive and separable form, the unitary evolution from these corrections is of the form $U = U_{1}\left({\bf x}\right) \otimes U_{2}\left({\bf y}\right)$ as required. 
This logic can be extended to show the factorized unitary nature of the time evolution to all orders in perturbation theory. Moreover, this logic can also be extended to field theory to show that the field theory evolution is also unitary. In perturbation theory, all that is necessary to establish the unitary nature of the evolution is to note at zeroth order in $\epsilon$, the time evolution is that of quantum field theory and is thus unitary. To first order in $\epsilon$, the non-linear correction enters as a Hermitean background classical field and thus the first order correction is also unitary. This can be extended to all higher orders in perturbation theory since the perturbative correction at any order relies solely on computing the lower order corrections which enter as classical background fields. 

We note that \cite{Polchinski:1990py} argued that the  non-linear evolution of separated systems must be additive in order to preserve causality. This is precisely the form of \eqref{eqn:separated} where we see that this structure naturally arises from field theory.\footnote{This form also violates one of the axions of \cite{Kapustin:2013yda} that proved a no-go theorem for modifications of quantum mechanics. \cite{Kapustin:2013yda} felt that the additive structure demanded by \cite{Polchinski:1990py} was not well motivated. We have demonstrated that this additive structure is a natural consequence of field theory.} The additive nature of non-linear evolution also implies that when a system is entangled with the environment, the environmental degrees of freedom are traced over in the non-linear evolution. This implies that non-linear effects persist even in the presence of decoherence. This was pointed out by \cite{Polchinski:1990py} and we develop the enormous experimental importance of this fact in Section \ref{sec:experiments}.

\subsection{Measurement}
\label{subsec:measurement}

Our definition of measurement parallels the operational concept of measurement described by the many worlds interpretation of quantum mechanics. In the many worlds interpretation, measurement is the time evolution of an initial state $|\Psi\rangle$ of some sub-systerm that is brought in contact with another sub-system (the measuring device) that is in a sufficiently stable initial state. This state could either be a stationary state of the sub-system or a coherent state that is robust against environmental decoherence. Label this initial state as $|A_0\rangle$. The system evolves as per the Schr\"{o}dinger equation in a deterministic way. In order to measure an observable associated with a Hermitian operator $\hat{O}$, whose eigenvectors are $|i\rangle$, the interaction Hamiltonian between the state and the measuring device must be such that the initial state $|\Psi\rangle\otimes|A_0\rangle$ time evolves into: 
\begin{equation}
\label{eqn:manyworldsQM}
|\Psi\rangle\otimes|A_0\rangle \rightarrow \sum_{i} c_i |i\rangle \otimes |A_i\rangle
\end{equation}
where $|\Psi\rangle = \sum_i c_i |i\rangle$. The $|A_i\rangle$ are a set of convenient ({\it i.e.} easily interpretable) states of the measuring device such as stationary or coherent states of the system.   When the $|A_i\rangle$ are stationary states, $\langle A_i |A_j\rangle = \delta_{ij}$. If the $|A_i\rangle$ are taken to be coherent states, the overlap is in general not zero - but for a good measuring device, the states are chosen so that the overlap is exponentially small.  Of course, all of this is a prediction of quantum mechanics and not an ``interpretation''. Due to the suppressed overlap between the $|A_i\rangle$, the reduced density matrix for the system  $|\Psi\rangle$ is effectively just $\sum_i |c_i|^2 |i\rangle\langle i|$. This reduced density matrix is interpreted classically as the direct sum of many possible outcomes $|i\rangle$ each with probability $|c_i|^2$ as determined by the states $|A_i\rangle$ of the measuring device. 

We adopt this operational procedure to define measurement: we bring the quantum state $|\Psi\rangle$ in contact with a measuring device that is initially in either a nearly stationary state or a coherent state ({\it i.e.} a state that is robust against environmental decoherence). Call this state as  $|\alpha_0\rangle$.  The interaction between these two systems will in general lead to an entangled state where the states of the measuring device are some desired states $|\alpha_j\rangle$ (such as nearly stationary states or coherent states). We will ``intepret'' this state in classical probabilistic terms. 
While the above describes the process of measurement, there are important differences in the phenomenology of measurement between linear quantum mechanics and this non-linear modification.

In linear quantum mechanics, the construction of the desired interaction between the system and the measuring device (the Hamiltonian) only requires knowledge of the eigenstates $|i\rangle$ of the operator $\hat{O}$ and the relevant states $|A_j\rangle$ of the measuring device. Knowledge of the actual state $|\Psi\rangle$ that is being measured is not necessary. In non-linear quantum mechanics the non-linear interactions depend on the actual quantum state $|\Psi\rangle$. There is thus a more complicated map between the outcomes of the measuring device and the incoming state unlike  the simple, direct map in linear quantum mechanics. 

But more importantly, unlike linear quantum mechanics, the states $|\alpha_i\rangle$ of the measuring device will in general overlap with each other at $\mathcal{O}(\epsilon)$. Normally (linear quantum mechanics) when one designs a measurement device, the time evolution is known and the Hamiltonian can be designed to result in orthogonal states.  Because the time evolution depends on the (unknown) initial state, one cannot necessarily guarantee this orthogonality.
Consequently, we expect $\langle \alpha_j | \alpha_i\rangle \propto \epsilon$.\footnote{ This statement can be easily verified for the stationary states of the system. It might be interesting to explicitly prove this for coherent states as well.} This implies that if the state $|\Psi\rangle\otimes|\alpha_0\rangle$ evolves into $\sum_i c_i |i\rangle|\alpha_i\rangle$, the reduced density matrix that describes $|\Psi\rangle$ will not be diagonal in the basis $|i\rangle$, and a partial trace over the measuring device subsystem gives $\text{Tr}_M (\rho) = \sum_{i, j} c_j^{*} c_{i} \langle \alpha_j | \alpha_i\rangle |i\rangle\langle j | \neq \sum_i |c_i|^2 |i\rangle\langle i| $. Consequently, even after measurement, interference between the states $|i\rangle$ and $|j\rangle$ can be observed. This state cannot be cleanly interpreted as the direct sum of many possible outcomes each with probability $|c_i|^2$ since the different worlds continue to interfere with each other. 

If one wishes to retain the clean interpretation of measurement as the direct sum of many classical worlds, the interaction Hamiltonian would have to evolve the initial state into $\sum c_i |i\rangle |A_i\rangle$ where the $|A_i\rangle$ are orthogonal states (for example, the usual energy eigenstates of the linear quantum mechanical hamiltonian governing the measuring device). But, these are not stationary states of the non-linear hamiltonian  and thus the measured outcome continues to evolve after the measurement. In this case, while the reduced density matrix of the quantum system itself has a ready classical interpretation as the direct sum of many outcomes, the time evolution of the measuring device does not reflect this clean interpretation. 

Further, in quantum mechanics once there is entanglement between the quantum state $|\Psi\rangle$ and a macroscopic measuring device, the system undergoes decoherence. Thus, different outcomes of the measurement cannot influence each other {\it i.e.} the world ``splits'' into many distinct worlds. As we will see in Section \ref{subsec:NLMacro}, non-linear effects can persist even in the presence of decoherence and thus the different outcomes or ``worlds'' can continue influencing each other. In general, these will also cause further time evolution of the states $|\alpha_j\rangle$ away from their values at the time of measurement. 

We thus learn that there isn't a clean concept of measurement in non-linear quantum mechanics where the state of the measuring device is faithfully correlated with the classical probabilistic reduced density matrix that describes the quantum system after measurement. It is thus the case that measurement in this non-linear system is unavoidably noisy. Our analysis of this non-linear system is perturbative and thus we would like to orient our discussion as close as possible to quantum mechanics. To that end, we will define measurement using the energy eigenstates (or coherent states)  $|A_i\rangle$ of the linear quantum mechanical Hamiltonian: measurement is a process that takes the quantum state $|\Psi\rangle \otimes |A_0\rangle \rightarrow \sum_i c_i |i\rangle |A_i\rangle$ with $|\Psi\rangle = \sum_i c_i |i\rangle$. This state can be interpreted as a classical direct sum of many possible outcomes $|i\rangle$ each with probability $c_{i}^{*}c_{i}/\langle \Psi| \Psi\rangle$ where $\langle \Psi| \Psi\rangle$ is the norm of the full quantum state.

\section{Analysis}
\label{sec:analysis}
In this section, we derive the key phenomenological consequences of this non-linear modification. This includes the  phenomenology of entangled macroscopic states (Subsection \ref{subsec:NLMacro}), the sensitivity of this effect to cosmology (Subsection \ref{subsec:cosmo}) and the causal aspects of this scenario in curved space-times (Subsection \ref{subsec:quantumwormholes}).

\subsection{Non-linearity and the Macroscopic World}
\label{subsec:NLMacro}

In quantum mechanics, it is difficult to observe quantum effects on macroscopic objects since the interactions of a macroscopic system with its environment leads to rapid decoherence. Upon decoherence, the macroscopic system is described by a classical probability distribution. The subsequent evolution of this system can be described purely in terms of evolving this classical probability distribution where each classical outcome evolves independently of the other outcomes. Interestingly, this is not the case for the non-linearities proposed here -  deviations from classical evolution are possible even in the presence of decoherence. 

We focus on the effects in non-relativistic quantum mechanics. The local nature of interactions implies that the position of a macroscopic body is rapidly entangled with the environment -- {\it i.e.}, macroscopic bodies are most easily described in the position basis. Suppose  $\Psi_M\left(\tau, {\bf x'}\right)$ is the wave function of a macroscopic body $M$ (where for notational simplicity, we are using a single position coordinate ${\bf x'}$ to represent the position coordinates of all the particles in the macroscopic body). Upon interaction with the environment $O$, this state evolves into an entangled state  whose wave-function is $\Psi_M\left(\tau, {\bf x'}\right)\Phi_O\left(\tau, {\bf y'};{\bf x'}\right)$ where $\Phi_O\left(\tau, {\bf y'};{\bf x'}\right)$ is the wave-function of the environment at time $\tau$ described by the position co-ordinate ${\bf y'}$ (again, for notational simplicity,  ${\bf y'}$  denotes the positions of all the particles in the environment). This wave-function is correlated with the position ${\bf x'}$ of $M$.  The non-linear evolution of $M$ is governed by the term 

\begin{equation*}
\int d^3{\bf x'} d^3{\bf y'} d\tau \Psi^{*}_{M}\left(\tau, {\bf x'}\right) \Psi_{M}\left(\tau, {\bf x'}\right)  \Phi^{*}_O\left(\tau, {\bf y'};{\bf x'}\right) \Phi_O\left(\tau, {\bf y'};{\bf x'}\right) G_R\left(t,{\bf x}; \tau, {\bf x'}\right)
\end{equation*}

The different environments entangled with different parts of the macroscopic body's wave function will generally be distinct enough to have essentially no overlap.  In other words, $\Phi^{*}_O\left(\tau, {\bf y'};{\bf x'}\right) \Phi_O\left(\tau, {\bf y'};{\bf x'}\right)$ will be `diagonal' when ${\bf y'}$ is integrated over.  
The integral thus inherits the time dependence of $\Psi^{*}_{M}\left(\tau, {\bf x'}\right)\Psi_{M}\left(\tau, {\bf x'}\right)$, {\it i.e.}, the actual time dependence of the full probability density of $M$. This probability density contributes as a time dependent potential that affects the evolution of $M$, leading to a deviation from classical evolution even in the presence of decoherence. The persistence of these effects even in the presence of decoherence  enables potent experimental probes of this scenario since the demands on environmental isolation of the probing systems can be significantly relaxed. We comment on these possibilities in Section \ref{sec:experiments}.

\subsection{Cosmological Sensitivity}
\label{subsec:cosmo}
The non-linear effects that we have described are tremendously sensitive to cosmic history - as we show below, even if these terms exist in the theory, many of their effects can become negligible under certain cosmic conditions such as canonical inflationary cosmology.

To see this, suppose the quantum state of the universe is such that it is in a macroscopic superposition $|\chi\rangle = \alpha |W\rangle + \beta |V\rangle$ of two states $|V\rangle$ and $|W\rangle$, where $|V\rangle$ is the vacuum state and $|W\rangle$ is the observer's world. The expectation value $\langle \chi| \phi| \chi \rangle = |\alpha|^2 \langle W|\phi|W \rangle$ and thus vanishes as $\alpha \rightarrow 0$, suppressing the effect of the non-linearity, independent of the value of $\epsilon$.  If the origin of the universe was such that the quantum state ended up in such a linear combination with $|\alpha| \ll 1$, there is no local operation that can be performed by the observer in $|W\rangle$ to boost the effects of the non-linearity. This is dramatically different from the effects of the standard linear evolution terms in quantum mechanics - their effects on the physics of $|W\rangle$ are independent of $\alpha$.  

The experimentally testable aspects of this theory are thus highly dependent on cosmology. We illustrate this by considering two scenarios, one where the universe's macroscopic structure was sensitive to early quantum fluctuations (Scenario A), and the other where the macroscopic structure was determined dominantly by classical evolution (Scenario B). Scenario A includes conventional inflationary cosmology in which the origin of structure in the universe is tied to quantum fluctuations of the inflaton field. These fluctuations decohere as they exit the horizon, giving rise to a quantum state that is the superposition of a very large number of distinct universes all of whom have the same statistical properties. This implies that the quantum state is such that in the vast majority of these universes, the space-time point where the Earth exists is likely to be in interstellar space. Thus, if a terrestrial observer tries to discover the effects of this non-linearity by trying to modify the value of the expectation value $\langle \chi | \phi | \chi \rangle$ (such as the experiments in Section \ref{subsec:classical}), the observer will have to contend with the tremendous suppression of this effect due to the small probability that the Earth exists at that specific space-time point (not to mention the observer themselves) in the full quantum state of the theory. The testable aspects of this theory would be along the lines described in Section \ref{subsec:quantum}. On the other hand, we could consider Scenario B - this could be an inflationary scenario where the perturbations have a classical origin, such as in warm inflation models \cite{Berghaus:2019whh} where the fluctuations are thermal. In this case, the evolution would be deterministic and the probability that a macroscopic object, such as the earth, is at this space-time point  could be non-negligible. The experiments described in Section \ref{subsec:classical} would then be a feasible path towards probing these effects. 

Importantly, while this cosmological sensitivity is transparent in this non-linear modification, it is a generic fact about any local non-linear modification of quantum mechanics. This is because the non-linear interactions, by the very nature of non-linearity, will depend upon the coefficients of the quantum state.  Local operations cannot undo small coefficients arising due to an unfavorable cosmic superposition. Importantly, in a local theory, one cannot boost these effects by writing down operators that would allow the observer to condition the effect based on the quantum state the observer finds themselves in. This is because there is no local operator that can represent the entire environment. Specifically, suppose the quantum state is $|\chi\rangle = \alpha_1 |O_1\rangle|E_1\rangle + \alpha_2 |O_2\rangle |E_2\rangle $ where the observer is in states $|O_{1, 2}\rangle$ corresponding to entanglement with the environmental states $E_{1, 2}$ respectively. If the non-linear terms governing the evolution of these terms were to be divided by a projection operator of the environmental state $E_1$, the observer $O_1$ could hope to see non-linear effects unsuppressed by $\alpha_1$. However, the projection on to the environmental state $E_1$ cannot be performed by a local operator since the environment, by definition, is a delocalized state.  

The generic nature of this cosmological sensitivity raises the interesting possibility that the observed linearity of quantum mechanics is the result of dynamical evolution of the universe (such as inflationary cosmology), even if the laws of quantum mechanics are fundamentally non-linear. 

\subsection{Curved Space}
\label{subsec:quantumwormholes}
In this section, we consider the possibility that the universe is placed in a linear superposition of metrics {\it i.e. } we consider quantum states of the form: 
\begin{equation}
\label{eqn:multiverse}
    |\chi\rangle = \alpha_1 |O_1\rangle|M_1\rangle + \alpha_2 |O_2\rangle|M_2\rangle
\end{equation}
 where the observers $O_{1,2}$ are entangled with two different metrics $M_{1,2}$. We clarify our formalism in this case and discuss possible new phenomena. We note in this non-linear theory,  the creation of such a superposition likely involves non-trivial gravitational dynamics since the matter in both these metrics can gravitationally influence each other. We will defer a detailed discussion of these dynamics for future work. In this section, we highlight phenomena that seem intuitively possible. 
 
An important aspect of non-linearity is that it removes the freedom to independently perform co-ordinate transformations on individual universes ({\it i.e. } metrics) $M_i$. There is still freedom to pick co-ordinates on the space-time manifold (general covariance) - but once these co-ordinates are picked, the co-ordinate change has to be reflected on all the parts of the wave-function. Unlike linear quantum mechanics where each metric in the superposition evolves independently, the non-linear evolution inextricably links all these metrics together. Non-linearity can thus be naturally embedded into the ${\bf 3} + {\bf 1}$ Hamiltonian formalism of General Relativity in which gravitation is described as the global time evolution of the metric on a spatial Cauchy surface. 

\subsubsection{Metrics and Contraction}
A generic field theoretic Lagrangian involves contraction between vectors and tensors that are performed using the metric tensor. We introduce non-linearity via expectation values of bosonic operators - in particular, these bosonic operators can be gauge fields (or really any tensor, including gravity). In linear quantum mechanics, since $\langle M_2 | M_1\rangle = 0$ there is never any ambiguity in contracting gauge fields to obtain the Lagrangian since the entanglement of the observer with a specific metric forces the observer to use that metric to contract tensors. In our case, the expectation value of the bosonic fields will get contributions from terms in \eqref{eqn:multiverse}.

 To see how to handle these terms, let us focus on the case of electromagnetism where the non-linear interaction is of the form $\epsilon g_{\mu \nu} \langle \chi | A^{\mu}| \chi \rangle J^{\nu}$. Formally, in this expression  $g_{\mu \nu}$ is an operator. Further $\langle \chi | A^{\mu}| \chi \rangle$ is simply a vector at a point on the manifold - this vector has an existence that is independent of any particular metric on the manifold. This term is then a well defined mathematical operation which makes the metric tensor operator $g_{\mu \nu}$ act on two vectors $\langle \chi | A^{\mu}| \chi \rangle$ and $J^{\nu}$. This operation is thus well defined. Now subject the state \eqref{eqn:multiverse} to non-linear evolution. Treating gravitation semi-classically, the observer $O_{i}$ will see the term $g_{\mu \nu}\left(|M_{i}\rangle\right) J^{\nu}\left(|M_{i}\rangle \right) \langle \chi | A^{\mu}| \chi \rangle $. Now, $\langle \chi | A^{\mu}| \chi \rangle = \alpha_{1}^{*}\alpha_1 \langle O_1, M_1 | A^{\mu}| O_1, M_1 \rangle +\alpha_{2}^{*}\alpha_2 \langle O_2, M_2 | A^{\mu}| O_2, M_2 \rangle $. These terms are generally described in terms of the co-ordinate systems used to describe the metrics $M_1$, $M_2$ respectively. But,  this  vector has a geometric meaning independent of the co-ordinate system on the manifold, and thus its components in any one co-ordinate system can be converted into those of any other co-ordinate system. In this non-linear case, this conversion is straightforward. A co-ordinate system was initially chosen on the space-time manifold in order to define the initial quantum state. While $M_1$ and $M_2$ were conventional co-ordinate charts used by $O_1$ and $O_2$, since they were constructed from the same underlying manifold, there is a natural map between these co-ordinate systems via the underlying co-ordinate chart on the space-time manifold. This map can be used to port vectors and tensors from one co-ordinate system to another.

\subsubsection{Quantum Wormholes}

Non-linear evolution coupled to metric superpositions allows for novel causal phenomena. To illustrate this, suppose the quantum state $|\chi\rangle$ is in the superposition \eqref{eqn:multiverse} with the state $|M_1\rangle$ being de Sitter space and $|M_2\rangle$ being Minkowski space. Pick two space-time points $x$ and $y$ such that they are not causally connected in $M_1$ but are causally connected in $M_2$. In linear quantum mechanics, $|M_1\rangle$ and $|M_2\rangle$ are two independent universes that do not communicate between each other. Accordingly, the observer $O_1$ cannot send signals from $x$ to $y$ while the observer $O_2$ can send signals between these two points. Non-linear evolution enables communication between these two ``worlds'' and allows for a quantum channel of communication. For example, in the Yukawa theory \eqref{eqn:Yukawa}, the observer $O_1$ can create an oscillating source of $\langle \phi \rangle$ using fermions in $M_1$ at $x$. This oscillating background can give rise to a propagating wave of $\phi$ in $M_2$. Since the space-time points $x$ and $y$ are causally connected in $M_2$, this wave can propagate in $M_2$ from $x$ to $y$. This propagating wave contains a non-zero $\langle \phi \rangle$ and it can thus influence fermions in $M_1$ at the point $y$. We call this phenomenon a ``quantum wormhole''. 

While this causal behavior is unusual, it is fundamentally causal - the quantum state of the universe has to allow the points to be causally connected in at least one of the metrics that are in superposition. The fact that causality on the manifold depends on the quantum state is not unusual - this is true even in linear quantum mechanics where it is possible that two points on a manifold are causally connected in one metric but not in another. The non-linear terms simply prevent a clear demarcation of the quantum  state into distinct universes/metrics. Note that this behavior does not require quantum non-linearities in gravitation - it exists even for the non-linear Yukawa theory as long as the quantum state itself is a superposition of different metrics.

It is interesting to ask if this kind of non-linear behavior could help solve the black hole information problem. The evaporation of a black hole will spread the position of the black hole over a distance $\sim r_s$ in a Page time. If there are non-linearities, the singularity that is behind the horizon in one metric may find itself in contact with a space-time point that is outside the horizon in another metric and that could in principle allow for information to escape from the singularity. While such communication is possible, this phenomenon alone cannot solve the black hole information problem - the non-linear terms are cosmologically sensitive. If the black holes were formed in a universe that was produced as a result of quantum inflationary perturbations, these non-linear terms will be vanishingly small and black holes formed in such a universe will not be able to lose their information via this mechanism.  But, in this case non-linear effects arising from the interference of metrics (see Sections \ref{subsubsec:metrics} and \ref{subsec:quantum}) can persist. It is possible that this metric interference would treat the event horizon as a special point (since one can no longer perform co-ordinate redefinitions without changing co-ordinates over all the universes) leading to new causal structures at the horizon. These causal structures may induce new vacuum energy divergences (similar to those found at the inner horizons of black holes -- see, for example, \cite{Hollands:2019whz} ) that could potentially result in the appearance of horizon scale singularities similar to  the firewall solutions discussed in \cite{Kaplan:2018dqx}. This possibility deserves further study.

Non-linearities  may also be useful as theoretical tools to define gauge invariant observables in gravitation. For example, if the quantum state was a superposition of a gravitating system and Minkowski space, the non-linear terms will allow the Minkowski observer to probe the physics of the gravitating system without this observer's measuring devices being affected by gravity. 

\subsubsection{Gravitational Phenomena From Interfering Metrics}
\label{subsubsec:metrics}

In this sub-section, we  speculate about  phenomena that may exist when quantum non-linearities are present in gravitation. The effective metric that is seen by an observer is then the sum $\tilde{g}_{\mu \nu}= \left(g_{\mu \nu } + \epsilon {\langle \chi|g_{\mu \nu}|\chi \rangle}\right)/(1+\epsilon)$ where $g_{\mu \nu}$ is the metric the observer experiences in the linear theory. With the non-linearity, the observer sees the effective metric $\tilde{g}_{\mu\nu}$ but would not be able to identify matter sources in his universe that would describe the dynamics arising from the term ${\langle \chi|g_{\mu \nu}|\chi \rangle}$. Using the linear theory, the observer would infer the existence of ``dark matter'' - which in this case would be the existence of a parallel universe. It would be interesting to examine this possible explanation for the observed dark matter in our universe - but we leave this investigation for future work. 

Novel gravitational dynamics may be possible in this situation. For example, in \eqref{eqn:multiverse} we may consider a quantum state that is a superposition of a crunching metric and a slowly expanding metric. In linear quantum mechanics the crunching metric would collapse to a singularity - but in this non-linear case, the contribution from the slowly expanding metric would alter the dynamics of the crunching state and may cause it to undergo a bounce and re-expand.  The observer in the crunching universe would view the contribution from the expanding metric as a background null energy violating source. But the source of this null energy violation is positive energy matter in the parallel universe. Similarly, metric interference from other universes may manifest itself observationally as a ``long distance modification'' of gravitation (see Section \ref{sec:pheno}) and may be tested with probes of such modifications. 

A key point to highlight in all of the above discussions is that the non-linear quantum terms are genuinely distinct from theoretical attempts to get novel gravitational phenomena by modifying gravity itself (for reviews, see \cite{Hinterbichler:2011tt,Levi:2018nxp}). In the latter, these modifications introduce additional degrees of freedom. In this non-linear modification, there are no additional degrees of freedom since the dynamical field content is identical to that of conventional linear quantum field theory.   The additional ``freedom'' arises purely from the freedom in the quantum states themselves.  While this freedom exists even in linear quantum mechanics, the non-linearity allows these states to influence their collective behavior. Note that unlike typical modifications of gravity, this theory does not appear to hit strong coupling at some arbitrary scale. The cutoff is the Planck scale.

\section{Constraints}
\label{sec:pheno}
 In this section, we discuss current experimental limits on these non-linearities. Owing to the cosmological sensitivity of non-linear quantum mechanics (see Section \ref{subsec:cosmo}), limits can only be placed within the context of a given cosmological scenario. We discuss constraints on three the non-linear couplings $\epsilon_{\gamma}$ and $\epsilon_{G}$ associated with non-linearities in electromagnetism and gravity respectively, which includes expectation values of their mediators and their sources.\footnote{Though we have not explicitly developed the theory, we could in principle explore bounds on $\epsilon_{S}$, non-linearity in the strong interactions.  For reasons discussed below, a naive reading of the expected effects of this coupling suggests no non-trivial bounds can be placed on this parameter.}  Many of the bounds we estimate can only be placed on this non-linear scenario in a classical universe where there is considerable overlap between the macroscopic observed world and the total wave-function of the universe. Even in this case, our analysis indicates that the bounds on these scenarios are rather weak but that they can be considerably improved with dedicated experiments. 
 
 Following this, we discuss constraints on the quantum scenario where our observed world has a tiny overlap  with the total wave-function of the universe. That is, we take the quantum state of the universe to be of the form $|\Psi\rangle = \frac{1}{\sqrt{N}} |U\rangle + \sqrt{1 - \frac{1}{N}}|R\rangle$ where $|U\rangle$ is our world and $|R\rangle$ represents vastly different worlds. We discuss two kinds of phenomena in this case. First, we consider expectation values sourced from our world. These expectation values are suppressed by $\frac{1}{N}$ and thus the quantity that affects non-linear evolution is the combination $\epsilon/N$. These vanish in the limit $N\rightarrow \infty$. Second, we consider the effects of the expectation value of the entire quantum state on physical processes in our universe. These are direct limits on $\epsilon$ but they are dependent on the state of the overall superposition. We focus on limits that exist even when $N \rightarrow \infty$ in this case {\it i.e.} limits that are resistant to quantum dilution. 

Naively, it would seem that violating quantum mechanics would dramatically alter pristine and well probed quantum systems such as atomic and nuclear states leading to significant constraints on $\epsilon_{\gamma}$ and $\epsilon_{S}$. This expectation fails due to the following reason.  In linear quantum mechanics, when confronted with a two body problem (such as an electron bound to a nucleus), one can simplify the problem by changing co-ordinates into the center of mass and relative co-ordinates. The center of mass motion of the system is irrelevant in determining the bound state energies of the system which only depend on the relative co-ordinates. In non-linear quantum mechanics, the entire wave-function of the particle is relevant. This is true not just in our specific modification but is a general property of (causal) non-linear systems. This implies that the center of mass motion of the system would enter in describing the effects of the non-linearity on the energy levels of the system. 

Let us analyze the specific case of the electromagnetic coupling $\epsilon_{\gamma}$ in the hydrogen atom. Suppose we localize the proton to essentially a fixed point in space -- we localize it so that the spread of the proton wave-function is much less than the Bohr radius $a_0 \sim 100$ pm of the atom. In this case, the non-linearity causes a self-interaction in the electron cloud and it can be verified that it will break the degeneracy between the 2S and 2P levels of Hydrogen and thus contribute to the Lamb shift. But, in a typical experiment that measures the Lamb shift the center of mass of the hydrogen atom is not localized. The spread in the center of mass suppresses the expectation value of the electromagnetic field which is ultimately the term responsible for shifting the energy levels in the atom due to the non-linearity. In a neutral atom that is localized to a distance $\lambda \gg a_0 $, the expectation value of the electromagnetic potential is non-zero only over a small thickness $\sim a_0$ at the edge of the region where the atom is localized. In typical measurements, the Lamb shift is measured via spectroscopy in a gas with vapor densities $\sim 10^{15} \text{ cm}^{-3}$ \cite{PhysRevLett.75.2470} yielding $\lambda \sim 0.1 \, \mu$m.

We estimate the maximum energy shift as follows\footnote{The actual limit requires detailed knowledge of the setup - we perform a conservative, order of magnitude estimate.}. Treat the hydrogen atom as being confined within a box of size $\lambda$. Let $\psi_0$ be the part of the total wave-function of the atom with the center of mass of the atom being in the bulk of the box and $\psi_1$ be the part where the center of mass of the atom being at the surface of the box within a thickness $\sim a_0$. The quantum state of the atom is of the form $\sim \psi_0 + \kappa \psi_1$ with $|\kappa|^2 \sim \frac{a_0}{\lambda} \sim 10^{-3}$. The electric field is essentially only non-zero in the surface region and is of order $\delta E \sim e/\lambda^2$. From perturbation theory, the energy shift is $\sim \epsilon_{\gamma} e |\kappa|^2 \langle \psi_1 | \delta E |\psi_1 \rangle \sim \epsilon_{\gamma} \alpha |\kappa|^2 \delta E a_0$. Requiring this energy shift to be smaller than the current uncertainty in the Lamb shift $\sim 5 \times 10^{-11}$ eV, we get $|\epsilon_{\gamma}| \lessapprox 10^{-4}$. This is a very conservative estimate since the atom is in fact not literally confined in a box of size $\lambda$ - the wave-function may spread beyond this, weakening this effect, and the boundaries of the region it is in may have canceling effects from other atoms. The Lamb shift has also been measured in ions \cite{PhysRevA.4.14} where the ions are also localized to within $\sim 0.1 \, \mu$m. The charge of the ion is spread over this distance $\lambda$ and the electric field from this charge can shift the energies of the electron. Requiring this electric field to be smaller than the background electric fields in these systems $\sim 50 \text{ V/cm}$ yields $|\epsilon_{\gamma}| \lessapprox 1$. 

A more stringent but sign-dependent bound on $\epsilon_{\gamma}$ can be placed by making use of the fact that traps have been used to trap ions. When $\epsilon_{\gamma}$ is positive, the non-linearity causes the ion's wave-function to repel itself leading to a repulsive potential $\sim \epsilon_{\gamma} \alpha_{EM}/L$ where $L$ is the size of the trap. This repulsive potential must be smaller than the confining potential of the trap. The confining potentials in the $L \sim 200$ nm traps used in \cite{PhysRevLett.75.4714} are $\sim$ 100 neV, yielding $\epsilon_{\gamma} \lessapprox 10^{-5}$. When $\epsilon_{\gamma}$ is negative, the non-linearity induces an attractive potential and thus the ion is more easily confined. A detailed mapping of the ion's confinement within the trap could constrain $\epsilon_{\gamma}$ in this case. It would be interesting to analyze data from current ion trap experiments to probe $\epsilon_{\gamma}$ in this case and we leave this possibility for future work. 

Similarly, in nuclear physics, it can be verified that the self interaction causes a state of total angular momentum $L$ to mix with states of total angular momentum $L \pm 2 n$ (where $n \subset \mathbb{Z}$) as long as $L \neq 0$. This mixing would lead to enhanced decay of isomers such as $^{180\text{m}}\text{Ta}$ that are stabilized by high angular momentum.  But, the decay rate due to non-linear effects is suppressed by the spread of the nuclear wave function in the gas or material and thus no useful limit can be extracted from the lifetime of such states.

The fact that non-linear quantum effects can persist in macroscopic bodies even in the presence of decoherence suggests that these superpositions would be the natural way to probe the gravitational non-linearity $\epsilon_G$. But, to realize this possibility, a macroscopic superposition needs to be created. There are straightforward ways of engineering such a superposition and, as we discuss in Section \ref{sec:experiments},  these can be used to experimentally probe $\epsilon_G$. But, we are not aware of any current experimental data from human-made systems that can be used to constrain $\epsilon_G$. It is also the case that the quantum spread of the wave-function of natural bodies whose gravitational effects are well understood are small - this is not a surprise since such bodies have a large mass and single quantum events, without the intervention of human engineering, do not back-react significantly on the positions of these macroscopic bodies. For example, a natural source of the spreading of the wave-functions of the Sun or the Earth is the radiation (thermal or particle emission) from these objects. We estimate that in the lifetime of the universe, these effects cause the wave-functions of the Sun and the Earth to spread by no more than $\lessapprox 10^{-11}$ km, a distance that is too small to be of observational importance. It thus appears that the only limits on $\epsilon_S$ and  $\epsilon_G$ are theoretical ones, requiring them to be smaller than  $\mathcal{O}\left(1\right)$ so that we may describe these systems using perturbative techniques and our extension is well-defined. 

All of the above limits on the various $\epsilon$ parameters are assuming a classically evolved universe ({\it i.e.}, $N=1$), but really should be considered bounds on $\et\sim \epsilon/N$.  In the quantum universe, $N\neq 1$, we first consider bounds that arise from sources in our universe. In this context, we can robustly constrain the quantity ${\epsilon_{\gamma}}/{N}$ from stellar energy loss arguments. The physical location of a star in our universe is likely to be a part of empty space in the other universes and thus time dependent expectation values of sources in the stellar interior can excite particles in other universes leading to energy loss. Using standard astrophysical results on stellar energy loss, we get $\frac{\epsilon_{\gamma}}{N}\lessapprox 10^{-13}$ \cite{Raffelt:2006cw}, thus rendering stronger bounds in this case than terrestrial ones. 

We now consider the case where the expectation value of the full quantum state affects the evolution of objects in our universe. When we have a large number $N \rightarrow \infty$ of universes in the full quantum state, we can only expect homogeneous effects to persist. Unlike any other quantum field, the metric is expected to have a non-vanishing expectation value in every quantum state. Moreover, since the total energy density in all these quantum states is positive, the expectation value of the metric across the entire quantum state will not average down to zero - it may, at worst get diluted to the non-zero flat space metric. A strikingly $N$-{\it independent} bound could potentially be placed on  $\epsilon_G$ in this scenario. This is because (for example) the Schwarzschild metric of an astronomical body will be polluted with a nearly locally flat metric from the rest of the wave function (where no spherical body exists). We defer a complete discussion of this possibility  for future work.

Note, these bounds on the quantum universe scenario are invalid when $N=1$ (classically evolved universe) because energy cannot be lost from the interior of a star to another part of the wave function if a star is located at the same place in every universe, and metrics are not changed for the same reason. 

\section{Experimental Opportunities}
\label{sec:experiments}

In this section, we highlight various experimental protocols and noise mitigation strategies that could be used to probe these non-linearities. In a companion paper, we present fully developed experimental proposals with sensitivity estimates. The cosmological sensitivity of the non-linearity necessitates different experimental strategies. Accordingly, we consider two possibilities - a classical universe where the observed world has significant overlap with the total wave function of the universe and an inflationary quantum universe where the universe is in a superposition of states which all have the same statistical properties.

Given the fact that quantum mechanics underlies the basic laws of nature, one may wonder if a classical universe is ever possible. Naively, one could expect quantum phenomena to result in runaway dynamical behavior that would place the universe in vastly different structures. But, there are good reasons to think that the world could be classical. While it is possible to engineer single quantum events to dramatically alter the behavior of macroscopic systems (as we propose to do in the following experimental protocols), there is reason to be skeptical that such events occur naturally. Suppose one needs to affect the state of $\sim n$ atoms to significantly alter the behavior of a macroscopic system. It is not unreasonable to expect $n \gtrapprox \mathcal{O}\left(10 - 100\right)$  (as opposed to $n \sim 1$). The macroscopic system is then governed by the average behavior of these $n$ atoms. The quantum mechanical random spread in this distribution is rapidly suppressed with $n$. The evolution is dominated by that of the expectation value which evolves in a deterministic and classical way. Interestingly, it is possible that $n \sim 1$ for biological systems and we comment about this possibility in Section \ref{sec:evolution}. 

The quantum cosmology case could be considerably more complicated than the inflationary scenario considered here with the superposition involving states that do not look anything like our universe at all. But some of the generic strategies outlined in the quantum case are likely to be useful in probing these broader scenarios as well.

\subsection{Classical Universe}
\label{subsec:classical}
In a classical universe, there is the potential to probe non-linear quantum mechanics via pristine atomic systems and through persistence of non-linear effects in macroscopic superpositions. We describe these in the following and then address the problem of ``quantum pollution'', a potentially worrying fact about  preserving these non-linearities given their fickle nature.

\subsubsection{Atomic  Systems}

The electromagnetic self-interaction term $\epsilon_{\gamma}$ can potentially be probed via single ion interferometry.  Take an ion and place it in a spatial superposition with the ion being at location $x_1$ with probability $p$ and $x_2$ with probability $1-p$. Now, hold the ion at those locations for a time $T$. The non-linear interaction will induce a relative phase shift between these two paths. Such a phase shift is absent in quantum mechanics. Importantly, this non-linear phase shift depends on the intensities $p$ and $1-p$ of the wave-functions at the two positions $x_1$ and $x_2$. This is unlike the case in quantum mechanics where the phase difference between two arms of an interferometer is independent of the intensity of the arms. In a single ion system, phase differences arise due to background noise. But, by using the fact that the non-linear signal depends on the intensity of the arms while the noise does not, it should be possible to engineer differential measurements that can robustly probe this effect. In fact, it is possible that data from current ion experiments could already place limits on these arm intensity dependent effects leading to bounds on $\epsilon_{\gamma}$ that are more stringent than the Lamb shift measurements described above. But, it is clear that a dedicated setup will offer a sharper probe. To obtain the maximum signal, it is advantageous to place the two arms of the ions as close as possible but without overlap in the respective probability densities. Note though that the estimated spread in the wavefunction of the Earth due to the emission of radiation limits the closest distance the ions can be brought together to around $\sim 10^{-11}$ km $= 10$ nm.  The signal also grows linearly with the interrogation time of the experiment.

\subsubsection{Macroscopic Superpositions}

The persistence of non-linear interactions even in the presence of decoherence enables the dramatic possibility of using macroscopic systems to test this theory. Consider the following protocol. Depending upon the outcome of a single quantum event (such as a spin measurement), we perform dramatically different macroscopic operations. In conventional quantum mechanics we will end up in different ``worlds'' with the respective outcomes. In each of those outcomes, we can now test for the existence of the ``other world'' by suitable sensors. In the following we outline a few experimental concepts to test various interactions. 

\begin{enumerate}
    \item Place a magnetometer next a coil of wire that initially has no current through it. Place the system in a macroscopic superposition by measuring the outcome of a spin measurement. For example, suppose we measure a spin-1/2 system. If we measure spin up, we turn on current into the coil. If we measure spin down, we leave the wire undisturbed. Then we measure the magnetic field in the magnetometer and see if there is an unexpected change in the magnetic field arising as a result of the current that was turned on in the ``other world''.  Electric systems in this vein may also be similarly constructed. This setup can be used to probe $\epsilon_{\gamma}$. This protocol could be implemented in an experimental setup such as CASPEr
    \cite{Budker:2013hfa}. 
    
    \item The gravitational coupling $\epsilon_G$ can potentially also be probed by using a protocol similar to the one proposed above. Consider an accelerometer, which could be an optical or atomic interferometer, and a large movable mass. Place the system in a superposition by a spin measurement. If we get spin up, we move the mass near the accelerometer. If we get spin down, we move it away from the accelerometer. In the case where we observed spin down, the accelerometer can be interrogated to probe the existence of the world where the spin was measured to be up.\footnote{An experiment along these lines was suggested in \cite{Page:1981aj} to see if gravitation was ``quantum'' or ``classical''. In our work, we highlight the fact that this is need not have a binary answer - a deformation of quantum mechanics permits all fundamental forces to have both quantum and classical behavior. } This protocol could be implemented in an experimental setup such as MAGIS  \cite{Graham:2012sy} or LIGO/VIRGO.
    
    \item In both of the systems considered above, it might also be possible to devise resonant systems that would boost the signal. If the measuring devices were resonant at a certain frequency, in all the cases where the spins were measured to be up, the macroscopic objects (currents, charges, masses) can be moved at the resonant frequency of the measuring device, amplifying the effect of the communication between these ``worlds''.
    
    \item To probe shorter range interactions such as QCD (and the weak interactions), one might choose to place material in the pathway of a beam depending upon the outcome of a spin measurement. One could look for the anomalous scattering of the beam in the world where the material was not placed in the beam's path, the origin of the scattering being the material that was along the beam's path in the ``other world''. These phenomena could be tested in beam dump experiments \cite{Izaguirre:2013uxa}.  These kinds of measurements can also be performed using coherent electromagnetic sources such as lasers and microwaves, for example, in experiments such as ALPS \cite{Ehret:2009sq}. 
\end{enumerate}

In all of the above cases, the experiments can be trivially modified to ensure that the effects are discovered in both parts of the macroscopic superposition. For example, in the case of the magnetometer experiment, we could have two well shielded coils with magnetometers in each of them. Depending upon the outcome of the spin measurement, the experimentalist can turn on the current in one coil and measure the magnet field in the other coil.  

\subsubsection{Evolutionary Dilution?}
\label{sec:evolution}
It is conceivable that while the large scale structure of the universe and the solar system are classical, significant quantum spread could have occurred in the evolution of biological systems. For example, it is conceivable that single quantum events may have had enormous impact on evolutionary dynamics, {\it e.g.}, the original formation or stability of RNA or mutations triggered by radioactive decays. In this scenario, it is possible that the formation of life on Earth has low probability and that in most of the wave function of the Universe, there is no life on Earth.  It is also possible that there are multiple biological civilizations that are currently co-existing on the Earth, all of them witnessing the same macroscopic classical universe. In these cases, the macroscopic effects discussed above will be suppressed due to the small overlap of the experiment with the rest of the wave-function. 

The following protocol could be used to detect non-linearities in the case of evolutionary dilution. One may construct a shielded room and place a radio telescope inside this room and use it to try and detect bright coherent astronomical radio sources. If evolution had lead to significant spread in local dynamics on the Earth, the shielded room would not be present at that spatial location in most of the wave-function. But, in a classical world, the astronomical source will be identical across the entire wave-function and thus lead to an unsuppressed expectation value of the electromagnetic field at that spatial point.  The non-linear coupling would allow the radio telescope to detect this expectation value without being suppressed by the local spread of the wave-function.  Another possible probe of this scenario is to try and measure the coherent magnetic field of the Earth inside a shielded room. The shield would not exist at the same spatial location in the rest of the wave-function, leading to a signal inside the shield. The fact that the GP-B experiment successfully created magnetic shielding at the level  $\sim$ 100 nG \cite{Everitt:2011hp} level implies a bound of $\epsilon_{\gamma} \lessapprox 10^{-7}$ in this scenario.

Interestingly, if such  effects were to be observed, not only would we have discovered non-linear quantum mechanics but we would also have unprecedented access to evolutionary dynamics. Intriguingly, in this case even though evolutionary dynamics dilutes the local non-linearity, it might be possible for human engineering to recover the full non-linear effect. One may for example consider game theory scenarios similar to those used by SETI to search for extra-terrestrial intelligence to send signals to other civilizations that may quantum mechanically co-exist on the Earth. If sufficiently many of them had also discovered non-linear quantum mechanics, it may be possible to communicate between these branches of the wave-function (using game theory scenarios, for example, using frequencies and locations of coherent astronomical sources) to coherently restore the possibility of exploiting quantum non-linearities. For example, a number of civilizations could conceivably agree to pick fixed locations on the earth to create coherent electromagnetic fields that may then be exploited as a community resource.

\subsubsection{Quantum Pollution}

The non-linearity is sourced by the expectation values of various operators and it is thus dependent on the behavior of the entire quantum state. Even in a hitherto classical world where single quantum events have not significantly changed the nature of the quantum state near the observer, there is always the potential that such changes could be caused due to the ease with which the world can be placed in macroscopic superpositions. These changes could dramatically suppress the ability to  detect and preserve quantum non-linearities. 

To illustrate this worry, suppose an experimentalist A performs the magnetometer experiment and discovers this effect. The results are announced and  experimentalist B  wants to test these results. Suppose it was the case that A's initial spin measurement was set to return spin up (down) with probability 0.5 (0.5). This implies that the universe is now in a quantum superposition. B is entangled with these two states of A. When B tries to repeat this experiment in B's laboratory,  unless both versions of B in this entangled state have current in their coil turned on at the same time, the effective magnetic field that could be detected by B would be lower than the corresponding effect seen by A. It is conceivable that these two entangled states of B may choose to turn on current in their coils at different times since it is not unreasonable that the discovery of such an effect could result in single quantum events dramatically altering macroscopic behavior even in an otherwise classical world. 
To mitigate this risk, we suggest the following protocols. 
\begin{enumerate}
    \item The initial macroscopic split by using the spin measurement need not be 0.5. It is likely prudent to sacrifice sensitivity and preserve the effect. One could then set up the initial spin measurement so that it returns spin up with a probability $p \ll 1$ and spin down with probability $1 - p$. In this case, the initial signal is smaller by $p$ and would thus require a more sensitive instrument to detect. However, this protects the non-linearity since the initial split can be repeated without significantly changing the wave-function. Indeed, for this reason, it would be wise to operate in the regime $p \ll 1$ even when A is trying to search for the non-linearity.
    \item It would be advantageous to run experiments for sufficiently long periods of time so that there is the improved possibility of overlap between the different versions of B. 
    \item The dilution of the non-linearity can also be decreased by communicating between different parts of the wave-function. For example, if A discovers these effects, A can use his setup as a way to communicate between different parts of the superposition and this communication channel can be used by B to ensure that both versions of B agree on experimental protocols. 
    \item In the best case scenario, the experiment is set up where both 'worlds' will see an effect -- for example, a spin measurements result in a macroscopic oscillating source of different frequencies.  The non-linear effects will allow detection of the oscillations at both frequencies, essentially the same result for both.  In addition, if the decision-making spin (quantum) measurement can automatically ({\it i.e.}, without active human intervention) trigger the macroscopic source, it may be possible for the experimentalist to see the results in an equivalent way in either world.
\end{enumerate}

\subsection{Inflationary Quantum Universe}
\label{subsec:quantum}
In canonical inflationary cosmology, the observed universe is in a macroscopic quantum superposition with a large number $N$ of other universes that all have the same statistical properties, but are locally completely different. Local experiments of the kind described in sub Section \ref{subsec:classical} are trying to measure the non-linearity by directly manipulating the expectation values of various fields. These experiments have to contend with two suppressions -  the non-linearity parameter $\epsilon$ and a suppression of $\sim 1/N$ from the fact that the experiment is only performed in a tiny part of the overall wave-function. In the large $N$ limit, these effects disappear, even if $\epsilon \sim {\cal O}(1)$. 

In this scenario, our observational strategy must hone in on the fact that in the vast majority of the wave-function, any space-time point, $x$, is in interstellar space, even though in our universe it may be inside a laboratory, the Earth, or a star. We can thus try to detect signatures that should be observable in interstellar space. Importantly, these observables need to source the expectation value of a bosonic field (such as electromagnetism) to be discoverable. The following signatures could be fruitfully pursued. 

\begin{enumerate}
    \item One could build a deep underground cosmic ray proton detector. The atmospheric and terrestrial overburden blocks cosmic ray protons from our universe from penetrating to this underground detector. But, the cosmic ray flux at the point $x$ in the wave-function is non-zero and this flux causes a time dependent electromagnetic field which can excite the underground detector. The IceCube experiment could potentially be used to search for this possibility but it would need to be able to distinguish high energy neutrino events from events caused by protons or millicharged particles. 
    \item One may similarly build a well shielded detector to search for coherent electromagnetic waves such as those produced by pulsars. These instruments would be similar to current dark matter detectors such as ADMX \cite{Khatiwada:2020mld} and DMRadio \cite{Chaudhuri:2014dla} that are looking for coherent dark matter waves. 
    \item One might hope for a signal from the dominant source of cosmic rays in the universe, namely the cosmic microwave background and starlight.  However, they do not give rise to these signatures. This is because in the quantum states of these particles $\langle \chi|A^{\mu}|\chi \rangle$ is effectively zero as these are thermal states. 
    
\end{enumerate}
In all of the above cases, the signal would average down with the number of universes as $\epsilon/\sqrt{N}$ because of their stochastic nature.  This is more favorably than the experiments described in Section \ref{subsec:classical}, but nevertheless vanishing in the large $N$ limit.

The reason that the above signatures average down as $1/\sqrt{N}$ is because of the fact that they are searching for effects that oscillate in time faster than the Hubble scale and are also varying in space. In a isotropic, homogeneous background these effects are expected to average down. We might thus expect isotropic and homogeneous effects that evolve over cosmological time scales to survive even when we have a large number of universes. This is indeed the case - for example, as we discussed earlier, quantum non-linearity in gravity would imply that an observer can detect metric interference in this universe. The observer would attribute this metric interference to a source of energy density that he is otherwise unable to detect {\it i.e.} some sort of ``dark energy''. But this ``dark energy'' would have an equation of state that tracks the cosmological energy density in the universe. Experiments that probe the equation of state of the universe may thus be well placed to probe these effects. Similarly metric interference also leads to departures from the Schwarzschild metric as discussed in Section \ref{sec:pheno} and manifests itself as a ``long distance modification'' of gravity. Tests of strong field General Relativity may thus be used to further probe this scenario. Moreover, as discussed in Section \ref{subsec:quantumwormholes} metric interference also likely leads to the creation of firewalls in black hole geometries and these may lead to signatures in gravitational wave detectors and the Event Horizon Telescope. We defer a detailed discussion of these effects to future work.

\section{Comparison with Prior Efforts} 
\label{sec:comparison}

Non-linear modifications to quantum mechanics have been considered in the past and elements of our approach have been discussed. We now describe the ways in which we have advanced these efforts. The specific framework for modifying quantum mechanics by incorporating state dependent expectation values in quantum field theory was considered by Kibble in \cite{Kibble:1978vm}. This work was largely focused on the ``measurement problem'' and pointed out that in scalar field theories such terms would give rise to state-dependent corrections to the mass of a particle. But, this work did not investigate the full scope of these modifications. Inspired by  \cite{Kibble:1978vm}, this class of modifications were further pursued by \cite{Stamp:2015vxa, Barvinsky:2018tsw, Carney:2018ofe, Barvinsky:2020pvw, Wilson-Gerow:2020moi}. These investigations focused on developing a gauge invariant path integral formalism to describe gravitation and discussed experimental consequences in the interference pattern of a single particle where quantum coherence of the interfering particle needs to be maintained. In our paper, we have shown that gauge invariance can be  incorporated in a straightforward manner for both gravitation and gauge theories. The experimental avenues that we pursue in this paper are focused on non-linear quantum effects that can be observed in macroscopic systems even in the presence of decoherence. 

Building on the formalism of Weinberg \cite{Weinberg:1989us}, Polchinski proposed \cite{Polchinski:1990py}  a causal, non-linear modification of single particle quantum mechanics. The persistence of non-linear quantum mechanical effects in the presence of macroscopic decoherence as well as the potential for the dilution of non-linear effects was pointed out in \cite{Polchinski:1990py}. In our work, the recognition that non-linear effects can be naturally incorporated in quantum field theory significantly changes the experimental approach towards probing these non-linearities. It also leads to a refined understanding of the dilution of non-linear effects where we recognize that this dilution is not inevitable but rather depends upon cosmic history. Further, we also recognize that there are some cosmological non-linear effects that are resistant to dilution. In the following, we elaborate on these aspects: 

\begin{enumerate}
    \item The investigations in \cite{Polchinski:1990py} were aimed at a general understanding of non-linear single particle quantum mechanics and thus the non-linear observables described were toy models that were single particle contact interactions. We have constructed explicit non-linear observables that arise from existing long range interactions in the theory such as electromagnetism and gravitation. These long range fields vastly change the experimental approach to detecting non-linear quantum mechanical effects: one may source long range fields from macroscopic systems ({\it e.g.}, the gravitational field of a macroscopic body) and use these to search for the non-linearity. This allows for an exciting experimental program since the effects could in principle be large as opposed to suppressed contact interactions. 
    
    \item The dilution of non-linearities is not automatic. Non-linear effects arise from the expectation values of quantum fields and as long as the large scale behavior of the universe is classical, all non-linear quantum effects can be experimentally accessed and exploited.  

    \item We point out that in canonical inflationary cosmology where quantum perturbations are the source of structure in the universe, many non-linear quantum effects can be diluted. The fact that non-linear effects can be associated with long range fields implies there are astronomical and cosmological signatures that can persist even in the presence of dilution (see sub-section \ref{subsec:quantum}). Indeed, some of the most dramatic effects such as metric interference leading to new effective sources of ``dark energy'' in the universe and the modification of the Schwarzschild metric persist even in the presence of extraordinary dilution (Section \ref{subsubsec:metrics}). 
    
    \item  In \cite{Polchinski:1990py},  the concept of measurement in non-linear quantum mechanics was discussed but not fully developed. We have developed this framework and pointed out that there is a consistent interpretation of measurement phenomena in non-linear quantum mechanics, albeit at the expense of accepting a fundamental source of error in all measurement processes.  
 
\end{enumerate}

\section{Conclusion}
\label{sec:concl}
In this paper, we have shown that field theory permits a natural way to introduce causal non-linear time evolution into quantum mechanics.   Surprisingly, despite the existence of pristine quantum environments such as atomic and nuclear systems, these modifications are presently ill-constrained. Moreover, we have also shown that any local non-linear modification of quantum mechanics is fundamentally fickle - it is highly sensitive to cosmic history and it also has the potential to dilute itself unless proper protocols are followed to preserve its effects. In addition, these non-linear effects are visible even when the quantum system decoheres. This makes it possible to test them in a variety of experiments, even when the underlying quantum state has a complex cosmic history. These are the key results of this paper. 

There are several avenues for continued exploration of these ideas. We have articulated many experimental protocols to test this scenario and it would be interesting to develop corresponding experimental proposals. These proposals should develop strategies to mitigate the potential fallout of the ``quantum pollution'' possibilities inherent in non-linear modifications. The sensitivity of this scenario to cosmic history makes these kinds of experiments especially important - a positive result would not just fundamentally overthrow the rules of quantum mechanics but it would also provide an unprecedented experimental probe into cosmic history. For example, a positive signal in an experiment that tests the ``classical universe'' scenario would show that the entire history of the universe has been deterministic. It would be a serious challenge to the conventional inflationary paradigm -  at the very least it would call for a classical source of perturbations that produced our universe. 

Alternately, if experimental measurements, such as those of the cosmic microwave background \cite{Green:2020whw}, prove the quantum origin of structure in our universe, it would dynamically explain the hitherto observed linear nature of  quantum mechanics.   It would also highlight the important role of cosmological measurements of the equation of state of the universe and tests of strong field General Relativity to search for non-linear effects that are resistant to such dilution. There is also a clear case to explore the role of non-linearities during inflation itself. If quantum non-linearities in either the inflaton or in gravitation are significant, it is possible that the many distinct universes that are typically produced in inflationary cosmology may significantly influence each other's evolution resulting in a very different quantum state than conventionally assumed. Smaller non-linearities could result in interesting non-gaussian structures where one might see a collision induced in the sky from a particle emerging from another universe, similar to the cosmic ray experiments proposed by us. 

The implications of discovering non-linearity in quantum mechanics are momentous. It allows for a rewrite of the fundamental rules of physics. We have seen that these non-linearities allow for entirely new causal behavior (such as quantum wormholes) and permit new phenomena in General Relativity (via metric interference) which may help solve the black hole information problem. It would also provide a useful theoretical tool to study strong gravity wherein one may conceive of observers who are co-located in a weakly gravitating parallel universe to probe the physics of strong gravity. The possibility that quantum mechanics itself could be non-linear raises important questions about the pursuit of the ultimate theory of nature that unifies quantum field theory with gravitation. These efforts are currently pursued based on the assumption that linear quantum mechanics holds to arbitrary energy. As we have seen, modifications of quantum mechanics can dramatically alter the behavior of physical systems and these modifications can be field dependent. Without experimental knowledge of these facts, it is difficult to see how purely mathematical pursuits of the ultimate theory of nature could result in a unique solution. 

There are remarkable technological implications as well: the key advantage offered by quantum computers is their ability to use superpositions to implement multiple computations in parallel.  In linear quantum mechanics, this advantage can be realized only if the system is able to retain quantum coherence despite the presence of the environment - a task that has proven to be challenging. The non-linear effects described by us would permit communication between different parts of this superposition. One may use a quantum event to place a classical computer into a superposition,  run a parallelized algorithm on the classical computer in these ``different worlds'' and then communicate the results of the computation using the non-linearity. While this parallelization realizes some of the benefits of a quantum computer, it does not achieve the full promise of a quantum computer. Unlike linear quantum mechanics, the non-linear effects decrease when the system is placed into a larger superposition. Thus, communication of the result of the computation across the entire wave-function requires more energy.  There is thus a trade-off between computation and the energy needed to communicate the results {\it i.e.} the power of the protocol increases with the number of times the wave-function is split but this also diminishes the overall strength of the non-linearity. The diminished strength can be compensated for by creating more energetic sources to communicate the non-linearity. This trade off between information and  energy is useful since the energy necessary for these purposes can be produced by brute force methods ({\it e.g.}, a power plant) as opposed to the cost associated with computing ({\it e.g.}, a supercomputer). But, the diminished strength of the non-linearity fundamentally implies that this protocol cannot solve NP problems without a prohibitive increase in the energy associated with communicating the result. This is unlike the case of linear quantum mechanics where the linear effects do not decrease as the system is placed in a large superposition. 

These benefits extend beyond just computing - non-linearities can also revolutionize quantum sensing. For example, the fundamental quantum limit on a sensor is set by shot noise. This shot noise limit can be vastly improved by exploiting the non-linearity. Suppose we have a spin 1/2 system where the spin is in the state $|S\rangle = \alpha |U\rangle + \beta |D\rangle$ and we want to know the coefficients $\alpha$ and $\beta$. We can now measure the spin in the bases $|U\rangle$ and $|D\rangle$ which would place the universe in the superposition $\alpha |U\rangle |M_U\rangle + \beta |D\rangle |M_D\rangle$, where $|M_{U, D}\rangle$ are states of the measuring device. In both universes, we now turn on a laser of fixed power and we now try to detect the laser from the other universe. Since the strength of the non-linearity depends on $\alpha$ and $\beta$, by measuring the available power, we infer $\alpha$ and $\beta$. Interestingly, this phenomenon permits the inference of the full quantum state with a single measurement as opposed to the repeated measurements necessary in linear quantum mechanics. It thus provides an independent realization of the ``Born rule''. 

Most stunningly, non-linearities would also make it possible  to parallelize large scale human efforts whose purpose is the discovery of information. For example, one may parallelize the search for natural resources. Suppose there is a natural resource in a large area. We may divide the area into a number of individual blocks. We can then appropriately split the wave-function of the universe. In each part of the wave-function we only search for the resource in a particular block. Upon discovery of the resource, the information can be transmitted across the rest of the wave-function using the non-linearity.   Given these mind blowing implications and the relatively straightforward experimental program that could be pursued to discover these effects, we believe there is a very strong case to explore this paradigm. 

\section*{Acknowledgments}

We thank Michael Fedderke, Daniel Green, Jason Hogan, Jared Kaplan,  Christian Spiering and Raman Sundrum for valuable discussions.  SR would like to thank R. Melmon for his friendship and  discussions about  quantum mechanics.

\bibliographystyle{unsrt}
\bibliography{references}

\begin{thebibliography}{10}

\bibitem{Bell:1964kc}
J.~S. Bell.
\newblock {On the Einstein-Podolsky-Rosen paradox}.
\newblock {\em Physics Physique Fizika}, 1:195--200, 1964.

\bibitem{Mermin:1998cg}
N.~David Mermin.
\newblock {What is quantum mechanics trying to tell us?}
\newblock {\em Am. J. Phys.}, 66:753, 1998.

\bibitem{Kochen:1968zz}
Simon Kochen and Ernst Specker.
\newblock {The Problem of Hidden Variables in Quantum Mechanics}.
\newblock {\em J. Math. Mech.}, 17:59--87, 1967.

\bibitem{Weinberg:1989us}
Steven Weinberg.
\newblock {Testing Quantum Mechanics}.
\newblock {\em Annals Phys.}, 194:336, 1989.

\bibitem{Polchinski:1990py}
Joseph Polchinski.
\newblock {Weinberg's nonlinear quantum mechanics and the EPR paradox}.
\newblock {\em Phys. Rev. Lett.}, 66:397--400, 1991.

\bibitem{GISIN19901}
N.~Gisin.
\newblock Weinberg's non-linear quantum mechanics and supraluminal
  communications.
\newblock {\em Physics Letters A}, 143(1):1--2, 1990.

\bibitem{Kibble:1978vm}
T.~W.~B. Kibble.
\newblock {Relativistic Models of Nonlinear Quantum Mechanics}.
\newblock {\em Commun. Math. Phys.}, 64:73--82, 1978.

\bibitem{Stamp:2015vxa}
P.~C.~E. Stamp.
\newblock {Rationale for a Correlated Worldline Theory of Quantum Gravity}.
\newblock {\em New J. Phys.}, 17(6):065017, 2015.

\bibitem{Barvinsky:2018tsw}
Andrei~O. Barvinsky, Daniel Carney, and Philip C.~E. Stamp.
\newblock {Structure of Correlated Worldline Theories of Quantum Gravity}.
\newblock {\em Phys. Rev. D}, 98(8):084052, 2018.

\bibitem{Carney:2018ofe}
Daniel Carney, Philip C.~E. Stamp, and Jacob~M. Taylor.
\newblock {Tabletop experiments for quantum gravity: a user\textquoteright{}s
  manual}.
\newblock {\em Class. Quant. Grav.}, 36(3):034001, 2019.

\bibitem{Barvinsky:2020pvw}
A.~O. Barvinsky, J.~Wilson-Gerow, and P.~C.~E. Stamp.
\newblock {Correlated Worldline theory: Structure and Consistency}.
\newblock {\em Phys. Rev. D}, 103(6):064028, 2021.

\bibitem{Wilson-Gerow:2020moi}
Jordan Wilson-Gerow and P.~C.~E. Stamp.
\newblock {Paths and States in the Correlated Worldline Theory of Quantum
  Gravity}.
\newblock 11 2020.

\bibitem{Kapustin:2013yda}
Anton Kapustin.
\newblock {Is there life beyond Quantum Mechanics?}
\newblock 3 2013.

\bibitem{Berghaus:2019whh}
Kim~V. Berghaus, Peter~W. Graham, and David~E. Kaplan.
\newblock {Minimal Warm Inflation}.
\newblock {\em JCAP}, 03:034, 2020.

\bibitem{Hollands:2019whz}
Stefan Hollands, Robert~M. Wald, and Jochen Zahn.
\newblock {Quantum instability of the Cauchy horizon in
  Reissner\textendash{}Nordstr\"om\textendash{}deSitter spacetime}.
\newblock {\em Class. Quant. Grav.}, 37(11):115009, 2020.

\bibitem{Kaplan:2018dqx}
David~E. Kaplan and Surjeet Rajendran.
\newblock {Firewalls in General Relativity}.
\newblock {\em Phys. Rev. D}, 99(4):044033, 2019.

\bibitem{Hinterbichler:2011tt}
Kurt Hinterbichler.
\newblock {Theoretical Aspects of Massive Gravity}.
\newblock {\em Rev. Mod. Phys.}, 84:671--710, 2012.

\bibitem{Levi:2018nxp}
Mich\`ele Levi.
\newblock {Effective Field Theories of Post-Newtonian Gravity: A comprehensive
  review}.
\newblock {\em Rept. Prog. Phys.}, 83(7):075901, 2020.

\bibitem{PhysRevLett.75.2470}
D.~J. Berkeland, E.~A. Hinds, and M.~G. Boshier.
\newblock Precise optical measurement of lamb shifts in atomic hydrogen.
\newblock {\em Phys. Rev. Lett.}, 75:2470--2473, Sep 1995.

\bibitem{PhysRevA.4.14}
Musti~A. Narasimham and Richard~L. Strombotne.
\newblock Lamb shift in singly ionized helium.
\newblock {\em Phys. Rev. A}, 4:14--32, Jul 1971.

\bibitem{PhysRevLett.75.4714}
C.~Monroe, D.~M. Meekhof, B.~E. King, W.~M. Itano, and D.~J. Wineland.
\newblock Demonstration of a fundamental quantum logic gate.
\newblock {\em Phys. Rev. Lett.}, 75:4714--4717, Dec 1995.

\bibitem{Raffelt:2006cw}
Georg~G. Raffelt.
\newblock {Astrophysical axion bounds}.
\newblock {\em Lect. Notes Phys.}, 741:51--71, 2008.

\bibitem{Budker:2013hfa}
Dmitry Budker, Peter~W. Graham, Micah Ledbetter, Surjeet Rajendran, and Alex
  Sushkov.
\newblock {Proposal for a Cosmic Axion Spin Precession Experiment (CASPEr)}.
\newblock {\em Phys. Rev. X}, 4(2):021030, 2014.

\bibitem{Page:1981aj}
Don~N. Page and C.~D. Geilker.
\newblock {Indirect Evidence for Quantum Gravity}.
\newblock {\em Phys. Rev. Lett.}, 47:979--982, 1981.

\bibitem{Graham:2012sy}
Peter~W. Graham, Jason~M. Hogan, Mark~A. Kasevich, and Surjeet Rajendran.
\newblock {A New Method for Gravitational Wave Detection with Atomic Sensors}.
\newblock {\em Phys. Rev. Lett.}, 110:171102, 2013.

\bibitem{Izaguirre:2013uxa}
Eder Izaguirre, Gordan Krnjaic, Philip Schuster, and Natalia Toro.
\newblock {New Electron Beam-Dump Experiments to Search for MeV to few-GeV Dark
  Matter}.
\newblock {\em Phys. Rev. D}, 88:114015, 2013.

\bibitem{Ehret:2009sq}
Klaus Ehret et~al.
\newblock {Resonant laser power build-up in ALPS: A
  'Light-shining-through-walls' experiment}.
\newblock {\em Nucl. Instrum. Meth. A}, 612:83--96, 2009.

\bibitem{Everitt:2011hp}
C.~W.~F. Everitt et~al.
\newblock {Gravity Probe B: Final Results of a Space Experiment to Test General
  Relativity}.
\newblock {\em Phys. Rev. Lett.}, 106:221101, 2011.

\bibitem{Khatiwada:2020mld}
R.~Khatiwada et~al.
\newblock {Axion Dark Matter eXperiment: Detailed Design and Operations}.
\newblock 9 2020.

\bibitem{Chaudhuri:2014dla}
Saptarshi Chaudhuri, Peter~W. Graham, Kent Irwin, Jeremy Mardon, Surjeet
  Rajendran, and Yue Zhao.
\newblock {Radio for hidden-photon dark matter detection}.
\newblock {\em Phys. Rev. D}, 92(7):075012, 2015.

\bibitem{Green:2020whw}
Daniel Green and Rafael~A. Porto.
\newblock {Signals of a Quantum Universe}.
\newblock {\em Phys. Rev. Lett.}, 124(25):251302, 2020.

\end{thebibliography}
\end{document}